\documentclass[11pt,a4paper]{article}
\usepackage{amssymb,amsmath,a4,pstricks,pst-plot}

\title{Tensor Product and Permutation Branes on the Torus}
\author{Cornelius Schmidt-Colinet\\[15pt] \textit{Institut f\"ur
    Theoretische Physik, ETH Z\"urich}\\\textit{CH-8093 Z\"urich,
    Switzerland}\\\textit{E-mail:} schmidtc@itp.phys.ethz.ch} \date{}

\begin{document}
\maketitle

\begin{abstract}
  We consider B-type D-branes in the Gepner model consisting of two
  minimal models at $k=2$. This Gepner model is mirror to a torus
  theory. We establish the dictionary identifying the B-type D-branes
  of the Gepner model with A-type Neumann and Dirichlet branes on the
  torus.
\end{abstract}

\section{Introduction}

D-branes in models with $N=(2,\,2)$ world-sheet supersymmetry have
been studied in various approaches and at different points in moduli
space, and it has been fruitful to combine several viewpoints (see
e.g.
\cite{Brunner:1999jq,Hori:2000ck,Diaconescu:2000ec,Mayr:2000as,%
Tomasiello:2000ym,Govindarajan:2001jc,Aspinwall:2006ib}).
In this paper we will study an example of the relationship between
D-branes in Gepner models (for some early work see
\cite{Recknagel:1997sb}), those of the corresponding geometric
compactification  
(see e.g.\ \cite{Ooguri:1996ck,Govindarajan:2000ef}),
and matrix factorisations of the equivalent Landau-Ginzburg theory 
that were first studied in 
\cite{Kapustin:2002bi,Brunner:2003dc,Hori:2004zd}.

There are two classes of branes that preserve half of the $N=2$
supersymmetry \cite{Warner:1995ay}; these are called A-type and
B-type, and are related by mirror symmetry. In the following we are
going to consider the B-type branes of a Gepner model involving two
minimal models at $k=2$, giving a total central charge of $c=3$. These
branes have an interpretation in terms of A-type branes in the
corresponding mirror, which is a torus theory \cite{Lerche:1989cs}.

We shall construct an explicit map between certain branes in the
Gepner model (tensor product and permutation branes
\cite{Recknagel:2002qq,Gaberdiel:2002jr}) and those of the torus,
matching the minimal model labels of the former with angles,
positions, and Wilson lines of the latter. This will be done by
writing the boundary states in either theory in terms of the Ishibashi
states of the diagonal $N=2$ theory at $c=3$. The Gepner model, on the
other hand, can be described topologically by an orbifold of the
Landau-Ginzburg theory with superpotential $W=x_1^4+x_2^4+z^2$ (see
e.g. \cite{Lerche:1989cs,Chun:1991js}), where the branes are described
by matrix factorisations of the superpotential. In a second step, we
shall identify the Gepner branes with matrix factorisations of the
Landau-Ginzburg theory \cite{Brunner:2005fv,Brunner:2005pq}.

A similar analysis has already been performed in
\cite{Gutperle:1998hb}. There the dictionary between the tensor
product Gepner branes and the torus branes was studied by comparing
the self-overlap of the boundary states. This leads to an
identification of the angle of the Gepner branes in the torus
description. Here we shall be more explicit; in particular we shall
also determine the relative positions and Wilson lines of the Gepner
branes, and we shall also discuss permutation branes. 

The relation between the tensor product branes of the Gepner model and
the branes of the LG theory with superpotential $W=x_1^4+x_2^4+z^2$
was also studied in  \cite{Govindarajan:1999js}. Finally, the 
branes of the LG theory with superpotential $W=x_1^4+x_2^4$, which is
mirror to the $\mathbb{Z}_4$ orbifold of the torus, were related in 
\cite{Dell'Aquila:2005jg} by matching intersection
matrices and the coupling to RR-primary fields.
\medskip

The organisation of the paper is as follows: In section
\ref{sec:torus}, we set up our notation for the torus theory and its
A-type D-branes; we also consider two $\mathbb{Z}_4$ symmetries whose
action on the branes has a geometric interpretation. 
In section \ref{sec:Gepner} the same is done for the corresponding 
Gepner model and its B-type tensor product and permutation branes. 
In particular we give the explicit formulae and propose two $\mathbb{Z}_4$ 
symmetries, one being the quantum symmetry of the orbifold, that 
correspond to those of the torus. Section \ref{sec:comparison} 
explains the matching of the branes on both sides, and in section 
\ref{sec:mf} we relate the Gepner branes to matrix factorisations 
of the corresponding superpotential. Section \ref{sec:conclusion} 
contains some conclusions.

\section{The torus $\mathcal{T}^2$}\label{sec:torus}

Let us begin by setting up our conventions for the conformal field
theory on the torus $\mathcal{T}^2$. The torus shall be rectangular,
and have both radii at the self-dual value. This theory is the mirror
of the Gepner model considered in section \ref{sec:Gepner}.

\subsection{Space of states}\label{ssec:torusstates}

The torus is given by two free bosonic fields
$X^1(z,\,\bar{z}),\,X^2(z,\,\bar{z})$, and two fermionic fields
$\psi^1(z)+\tilde{\psi}^1(\bar{z}),\,\psi^2(z)+\tilde{\psi}^2(\bar{z})$,
where we have explicitly written out the chiral and antichiral parts.
The bosonic fields are each compactified on a circle of self-dual
radius $R=1$ (for $\alpha '=1$):
 \[
 X^i(z,\bar{z})\sim X^i(z,\bar{z})+2\pi\qquad(i=1,\,2)\,.
 \]
We can complexify these fields as
\begin{displaymath}
X^{\pm}:=\frac{1}{\sqrt{2}}(X^1\pm
iX^2)\,,\quad\psi^{\pm}:=\frac{1}{\sqrt{2}}(\psi^1\pm
i\psi^2)\,,\,\quad\tilde{\psi}^{\pm}:=\frac{1}{\sqrt{2}}(\tilde{\psi}^1\pm
i\tilde{\psi}^2)\,.
\end{displaymath}
For the bosonic fields, the derivatives with respect to $z$
($\bar{z}$) are chiral (antichiral) fields with mode expansion
\begin{equation}\label{Xmodes}
\partial_z
X^{\pm}=-i\sum_{n\in\mathbb{Z}}\alpha^{\pm}_n\,z^{-n-1}\,,
\quad\partial_{\bar{z}} 
X^{\pm}=-i\sum_{n\in\mathbb{Z}}\tilde{\alpha}_n^{\pm}\,\bar{z}^{-n-1}\,;
\end{equation}
the chiral fermionic fields have the expansion
\begin{equation}\label{psimodes}
\psi^{\pm}=\sum_{r}\psi^{\pm}_r\,z^{-r-\frac{1}{2}}\,,\quad
\tilde{\psi}^{\pm}=\sum_{r}\tilde{\psi}^{\pm}_r\,\bar{z}^{-r-\frac{1}{2}}\,,
\end{equation}
where $r\in\mathbb{Z}$ in the Ramond sector and
$r\in\mathbb{Z}+\frac{1}{2}$ in the Neveu-Schwarz sector. Due to the
compactification, the ground states in the space of states have
momenta given by momentum and winding quantum numbers $p_i$ and
$w_i$,
\begin{equation}\label{ground_state_momenta}
P^L_i=\tfrac{1}{\sqrt{2}}(p_i+w_i),\quad
P^R_i=\tfrac{1}{\sqrt{2}}(p_i-w_i)\quad(i=1,\,2)\,.
\end{equation}
The superscripts $L$ and $R$ of the center of mass momenta $P$ refer
to left- and right-moving fields.\\ 

We will be interested in the $N=2$ supersymmetry of this theory.  The
Verma module with respect to the $N=2$ generators on each of the
ground states (\ref{ground_state_momenta}), except for the vacuum
state $p_i=w_i=0$, forms an irreducible $N=2$ highest weight
representation at $c=3$. In the NS sector, the corresponding highest
weight state has conformal dimension $H$ ($\tilde{H}$) and $U(1)$
charge $Q$ ($\tilde{Q}$) for the left-(right-)movers, with
\begin{equation}\label{torushighestweights}
\begin{array}{ll}
H=\frac{1}{4}((p_1+w_1)^2+(p_2+w_2)^2)\,,& \quad Q=0\,,\\[4pt]
\tilde{H}=\frac{1}{4}((p_1-w_1)^2+(p_2-w_2)^2)\,,& \quad \tilde{Q}=0\,.
\end{array}
\end{equation}
Highest weights and charges of the R sector states are reached by
spectral flow, which gives rise to a representation at highest weight
$H+\tfrac{1}{8}$ for every NS representation at highest weight
$H>0$. We will use the convention that  we label a Ramond
representation by conformal dimension and charge of the highest weight
vector which is annihilated by the mode $G_0^+$.\\ 
Ground states with momenta as in (\ref{ground_state_momenta}) will be
denoted
\begin{equation}\label{groundstates}
|p_1,\,w_1,\,p_2,\,w_2\rangle_{\mathrm{NS,\,R}}
\end{equation} 
for momentum quantum numbers $p_i$ and winding numbers $w_i$. We will
drop the R or NS index when unnecessary.\\ 

The Verma modules built on the vacuum states $p_i=w_i=0$ are reducible
in both the NS and the R sector, as in the uncompactified case
\cite{Gaberdiel:2004nv}. In the NS sector, highest weights and charges
of these representations are given by 
\begin{equation}
\begin{array}{lll}
H=0\,,&\quad Q=0\,;&\\[4pt]
H=\frac{2|n|-1}{2}\,,&\quad Q=\mathrm{sign}(n) &\quad
\mathrm{for}\;n\in\mathbb{Z}\setminus \{0\}\,.
\end{array}
\end{equation}
The R sector representations follow again with the help of the
spectral flow. 

These representations are generated in the NS sector from the singular
vectors of the $N=2$ vacuum Verma module 
\begin{equation}\label{singularNS}
\begin{array}{l}
(\alpha^+_{-1})^{n-1}\psi^+_{-\frac{1}{2}}|0,\,0,\,0,\,0\rangle_{\mathrm{NS}}
\,,\\[5pt] 
(\alpha^-_{-1})^{n-1}\psi^-_{-\frac{1}{2}}|0,\,0,\,0,\,0\rangle_{\mathrm{NS}}\,,
\end{array}
\end{equation}
for $n\in\mathbb{N}$; we will use the short-hand notation
\begin{equation}\label{vacstatesNS}
|n\rangle_{\mathrm{NS}}=\left\{
\begin{array}{lr}
|\frac{2|n|-1}{2},\,1\rangle_{\mathrm{NS}}&\mathrm{for\;}n>0\,,\\[4pt]
|0,\,0\rangle_{\mathrm{NS}}&\mathrm{for\;}n=0\,,\\[4pt]
|\frac{2|n|-1}{2},\,-1\rangle_{\mathrm{NS}}&\mathrm{for\;}n<0\,,
\end{array}\right.
\end{equation}
where the right-hand side gives the conformal dimension and charge of
the corresponding highest weight vector. Here, the states with $n>0$
denote the states of the first line in (\ref{singularNS}), and those
with $n<0$ the states of the second line in (\ref{singularNS}). 

In the R sector, the singular vectors are
\begin{equation}\label{singularR}
\begin{array}{ll}
(\alpha^+_{-1})^{n-1}\psi^+_{-1}|0,\,0,\,0,\,0\rangle_{\mathrm{R}}\,,&
\quad (\alpha^+_{-1})^{n}|0,\,0,\,0,\,0\rangle_{\mathrm{R}}\,;\\[6pt]
(\alpha^-_{-1})^{n-1}\psi^-_{-1}\psi_0^-|0,\,0,\,0,\,0\rangle_{\mathrm{R}}\,,&
\quad (\alpha^-_{-1})^{n}\psi_0^-|0,\,0,\,0,\,0\rangle_{\mathrm{R}}\,,
\end{array}
\end{equation}
where $|0,\,0,\,0,\,0\rangle_{\mathrm{R}}$ is the free field 
ground state $|\tfrac{1}{8},\tfrac{1}{2}\rangle_{\mathrm{R}}$. 
We will use a short-hand notation analogous to the NS case, namely
\begin{equation}\label{vacstatesR}
\begin{array}{l}
|n,+\rangle_{\mathrm{R}}=\left\{\begin{array}{ll}|\frac{1}{8},\,
\frac{1}{2}\rangle_{\mathrm{R}}&(n=0)\\[3pt]|n+\tfrac{1}{8},\,
\tfrac{3}{2}\rangle_{\mathrm{R}}&(n\in\mathbb{N})\end{array}\right.
\,,\\[16pt] 
|n,-\rangle_{\mathrm{R}}=\left\{\begin{array}{ll}|\frac{1}{8},\,
-\frac{1}{2}\rangle_{\mathrm{R}}&(n=0)\\[3pt]|n+\tfrac{1}{8},\,
-\tfrac{3}{2}\rangle_{\mathrm{R}}&(n\in\mathbb{N})\end{array}\right.\,.
\end{array}
\end{equation}

\subsection{Two $\mathbb{Z}_4$ symmetries on
  $\mathcal{T}^2$}\label{ssec:torusZ_4} 

We note two $\mathbb{Z}_4$ symmetries that we will identify in chapter
\ref{sec:comparison} with symmetries of the corresponding Gepner
model. 

The rotation group $\mathbb{Z}_4$ acts naturally
on the bosonic torus fields when the action of its generator $g$ on
the fields is given by
\begin{eqnarray}\label{naturalZ_4}
&&g(X^1)=-X^2 \,,\qquad g(X^2)=X^1\,, \nonumber \\
\\[-15pt]
&&g(\psi^1)=-\psi^2 \,,\qquad g(\psi^2)=\psi^1\,, \nonumber
\end{eqnarray}
which in terms of the complexified fields reads
\begin{displaymath}
g(X^{\pm})=e^{\pm i \frac{\pi}{2}}X^{\pm}\,,\qquad g(\psi^{\pm})
=e^{\pm i \frac{\pi}{2}}\psi^{\pm}\,.
\end{displaymath}
A little care is required when we define the phase of the action of
$g$ on the ground states with non-vanishing momentum
(\ref{groundstates}). In the NS sector, we can define 
\begin{equation}\label{gaction}
g|p_1,\,w_1,\,p_2,\,w_2\rangle_{\mathrm{NS}}=|-p_2,\,-w_2,\,p_1,\,
w_1\rangle_{\mathrm{NS}}\,.
\end{equation}
In the R sector, where the ground states form a tensor product of two
two-dimensional representations of the Dirac algebra, we must include
an appropriate phase.

The highest weight states in the vacuum sectors obtain a phase under
the $\mathbb{Z}_4$ action according to (\ref{singularNS}),
(\ref{singularR}): 
\begin{equation}\label{Z_4torusvacuum}
\begin{array}{r@{\;=\;}ll}
g|n\rangle_{\mathrm{NS}}&e^{i\pi n}|n\rangle_{\mathrm{NS}}&
\qquad(n\in\mathbb{Z})\,,\\[4pt]
g|n,\,\pm\rangle_{\mathrm{R}}&e^{\pm i\pi (n+\frac{1}{2})}|n,\,
\pm\rangle_{\mathrm{R}}&\qquad(n\in\mathbb{N}_0)\,.
\end{array}
\end{equation}
A linear combination of ground states which is an eigenstate of
eigenvalue $e^{i\frac{\pi}{2}t}$ for $0\leq t\leq 3$ with respect to
this $\mathbb{Z}_4$ symmetry will be denoted with a superscript $t$: 
\begin{equation}\label{Z_4eigenstate}
|p_1,\,w_1,\,p_2,\,w_2\rangle^t=\frac{1}{2}
\displaystyle{\sum_{n=0}^3}e^{-i\frac{\pi}{2}tn}
g^n|p_1,\,w_1,\,p_2,\,w_2\rangle\,.
\end{equation}

The other symmetry is a $\mathbb{Z}_4$ symmetry involving T-duality,
which we will call $\mathbb{Z}_4'$ in order to distinguish it from the
previous one. Denoting its generator $g'$, it acts on the ground
states (\ref{groundstates}) as 
\begin{equation}\label{g'action}
g'|p_1,\,w_1,\,p_2,\,w_2\rangle_{\mathrm{NS}}=
(-1)^{p_1+p_2}|w_1,\,p_1,\,w_2,\,p_2\rangle_{\mathrm{NS}}\,. 
\end{equation}
In the vacuum sectors, this symmetry has the same effect as
(\ref{Z_4torusvacuum}).

An eigenstate with respect to both symmetries is denoted 
\begin{equation}\label{allZ_4eigenstate}
|p_1,\,w_1,\,p_2,\,w_2\rangle^{t,m}=\frac{1}{4}
\sum_{n=0}^{3}\sum_{s=0}^{3}e^{-i\frac{\pi}{2}(st+mn)}
g^s(g')^n|p_1,\,w_1,\,p_2,\,w_2\rangle\,.
\end{equation}
The superscript on the left-hand side indicates the eigenvalues
$e^{i\frac{\pi}{2}t}$ under $g$ and $e^{i\frac{\pi}{2}m}$ under
$g'$.\\ 

The first $\mathbb{Z}_4$ action can be interpreted geometrically
as a rotation by 90 degrees. The action of $\mathbb{Z}'_4$ amounts to
a T-duality transformation in both directions, and the phase can be 
seen as a shift $X^i(z,\bar{z})\mapsto X^i(z,\bar{z})+\pi$ in
both directions, i.e. as
$X_{\mathrm{L,R}}^i\mapsto X^i_{\mathrm{L,R}}+\tfrac{\pi}{2}$.

\subsection{The $N=2$ boundary states on
  $\mathcal{T}^2$}\label{ssec:torusbranes} 

We are interested in boundary states on the torus that satisfy the
$N=2$ boundary conditions of type A, 
\begin{eqnarray}\label{sconfbcA}
(L_n-\tilde{L}_{-n})\,\|A\rangle\rangle&=&0\,,\nonumber\\
(J_n-\tilde{J}_{-n})\,\|A\rangle\rangle&=&0\,,\\
(G_r^{\pm}+i\eta \tilde{G}_{-r}^{\mp})\,\|A\rangle\rangle&=&0\,,\nonumber
\end{eqnarray}
with spin structure $\eta\in\{\pm1\}$. The zero mode condition is
$H=\tilde{H}$ and $Q=\tilde{Q}$, which means  
\begin{equation}\label{ishibashicondition}
p_1w_1=-p_2w_2\,
\end{equation}
in terms of the ground state quantum numbers. For non-vanishing momenta, 
there is -- up to a phase -- a unique Ishibashi state in these 
representations, which we will denote by \cite{Ishibashi:1988kg,Klemm:2005dt}
\begin{equation}\label{ishibashistateoldbasis}
|p_1,\,w_1,\,p_2,\,w_2;\,\eta\rangle\rangle_{\mathrm{NS,\,R}}\,.
\end{equation}
The subscript is to be understood as specifying either the NS-NS or the 
R-R sector. We can fix the relative normalisations between Ishibashi 
states at given highest weight by demanding that these states transform 
under the $\mathbb{Z}_4$ symmetries in the same way (\ref{gaction}), 
(\ref{g'action}) as the NS ground states, i.e. by setting
\begin{equation}
\begin{array}{r@{\;=\;}l}
g|p_1,\,w_1,\,p_2,\,w_2\rangle\rangle_{\mathrm{NS,\,R}}&
|-p_2,\,-w_2,\,p_1,\,w_1\rangle\rangle_{\mathrm{NS,\,R}}\,,\\[4pt]
g'|p_1,\,w_1,\,p_2,\,w_2\rangle\rangle_{\mathrm{NS,\,R}}&(-1)^{p_1+p_2}
|w_1,\,p_1,\,w_2,\,p_2\rangle\rangle_{\mathrm{NS,\,R}}\,.
\end{array}
\end{equation}

In the vacuum sector, the representations containing the same singular 
vectors in the left- and the right-moving part of the theory are 
isomorphic, so that we have an Ishibashi state for every left-moving 
irreducible highest weight representation. We will denote the Ishibashi 
states in the vacuum sectors analogously to the left-moving ground 
states by
\begin{equation}
|n;\,\eta\rangle\rangle_{\mathrm{NS}},\qquad|n,\,\pm;\,\eta
\rangle\rangle_{\mathrm{R}},
\end{equation}
where $n\in\mathbb{Z}$ in the NS-NS and $n\in\mathbb{N}_0$ 
in the R-R sector.\\

It was shown in \cite{Gaberdiel:2004nv} that the $N=2$ boundary states on 
the torus can all be expressed in terms of the usual Neumann branes with 
electric fields. The Neumann gluing conditions with flux $\phi$,
\begin{equation}\label{torusgluingconditions}
\begin{array}{r@{\,=\,}l}
(\alpha^{\pm}_n+e^{\mp i\phi}\,\tilde{\alpha}^{\mp}_{-n})\,\|A
\rangle\rangle&0\,,\\[4pt]
(\psi^{\pm}_r+i\eta e^{\mp i\phi}\,\tilde{\psi}^{\mp}_{-r})\,\|A
\rangle\rangle&0\,,
\end{array}
\end{equation}
imply the $N=2$ gluing conditions (\ref{sconfbcA}) for every (real) 
value of $\phi$. On the other hand, every fundamental $N=2$ boundary 
state on the torus is actually a state satisfying 
(\ref{torusgluingconditions}) for a specific flux $\phi$. 
In fact, for non-vanishing ground-state momentum, any $N=2$ Ishibashi 
state $|p_1,\,w_1,\,p_2,\,w_2;\,\eta\rangle\rangle$ actually defines 
a $U(1)$ Ishibashi state satisfying (\ref{torusgluingconditions}), 
with $\phi$ given by
\begin{equation}\label{torusangle}
\tan \left(\frac{\phi}{2}\right)=\frac{p_1}{p_2}\,.
\end{equation}
This expression allows us to interpret the quantity $\phi$ as an 
angle in the space of momentum quantum numbers. 
In the NS sector, a convenient notation is to label the $U(1)$ 
states by two coprime momentum quantum numbers 
$\hat{p}_1\in \mathbb{Z},\,\hat{p}_2\in\mathbb{N}_0$, and two integers 
$a,\,b$,
\begin{equation}\label{ishibashistatenormalbasisNS}
|a,\,b,\,\phi;\,\eta\rangle\rangle_{\mathrm{NS}}:= |a\hat{p}_1,\,
b\hat{p}_2,\,a\hat{p}_2,\,-b\hat{p}_1;\,\eta\rangle\rangle_{\mathrm{NS}}\,,
\end{equation}
where the right-hand side is in the notation (\ref{ishibashistateoldbasis}),
 and $\phi$ satisfies (\ref{torusangle}), i.e. 
$\tan (\phi/2)=\hat{p}_1/\hat{p}_2$. 

In the R sector, we have to be more careful. 
It turns out that we can avoid the action of $g$ to look quite tedious 
in both sectors by extending the angle $\phi$ to take values in the interval 
$(-2\pi,2\pi]$, i.e. to enlarge its period to $4\pi$. We will hence 
associate to every Ishibashi state in the representation labelled by 
$p_1=a\hat{p}_1,\,w_1=b\hat{p_2},\,p_2=a\hat{p}_2,\,w_2=-b\hat{p}_1$ the 
two notations
\begin{equation}
|a,\,b,\,\phi\rangle\rangle_{\mathrm{R}}=-|-a,\,-b,\,\phi+2\pi
\rangle\rangle_{\mathrm{R}}\,,
\end{equation}
and do the same in the NS sector, but without the relative minus sign. 
Arranging the signs, the action of $g$ and $g'$ on the Ishibashi states 
in the new notation can be written as
\begin{eqnarray}
g|a,\,b,\,\phi\rangle\rangle&=&|a,\,b,\,\phi+\pi\rangle\rangle\,,
\nonumber\\
\\[-10pt]
g'|a,\,b,\,\phi\rangle\rangle&=&|-b,\,a,\,\phi+\pi\rangle\rangle\nonumber
\end{eqnarray}
in both the R and the NS sector.

In the vacuum sectors, the $N=2$ Ishibashi states transform analogously 
to their respective ground states, i.e.
\begin{eqnarray}\label{Z_4torusvacuumishibashis}
\begin{array}{r@{\;=\;}ll}
g|n;\,\eta\rangle\rangle_{\mathrm{NS}}&e^{i\pi n}|n;\,\eta\rangle
\rangle_{\mathrm{NS}}&\qquad(n\in\mathbb{Z})\,\\[5pt]
g|n,\,\pm;\,\eta\rangle\rangle_{\mathrm{R}}&e^{\pm i\pi 
(n+\frac{1}{2})}|n,\,\pm;\,\eta\rangle\rangle_{\mathrm{R}}&
\qquad(n\in\mathbb{N}_0)\,.
\end{array}
\end{eqnarray}
We will now fix the remaining phases in the definition of our Ishibashi 
states. In the vacuum sectors, this is done by setting
\begin{eqnarray}
\|\mathrm{Neumann};\,\eta\rangle\rangle_{\mathrm{NS}} &=& \mathcal{N}
\sum_{n\in\mathbb{Z}}|n;\,\eta\rangle\rangle_{\mathrm{NS}}\qquad
\mathrm{and}\nonumber\\
\|\mathrm{Neumann};\,\eta\rangle\rangle_{\mathrm{R}} &=& \mathcal{N}
\sum_{n\in\mathbb{N}_0}\bigg(|n,\,+;\,\eta\rangle
\rangle_{\mathrm{R}}+|n,\,-;\,\eta\rangle\rangle_{\mathrm{R}}\bigg)\,,
\nonumber
\end{eqnarray}
where $\|\mathrm{Neumann};\,\eta\rangle\rangle_{\mathrm{NS},\,\mathrm{R}}$ 
is the free field Neumann vacuum boundary state. The phases of the other 
Ishibashi states are determined by writing the boundary states in the 
following form \cite{Klemm:2005dt}:
\begin{equation}\label{torusbrane}
\begin{array}{r@{}l}
\|A,\,B,\,\phi,\epsilon;\,\eta\rangle\rangle=\mathcal{N}(\phi)\bigg\{&
\displaystyle{\sum_{n\in\mathbb{Z}}e^{in\phi}\,|n;\,\eta\rangle
\rangle_{\mathrm{NS}}}\\[4pt]
&+\displaystyle{\sum}_{(a,b)\in\mathbb{Z}^2\setminus\{(0,0)\}}e^{iAa+iBb}
|a,\,b,\,\phi;\,\eta\rangle\rangle_{\mathrm{NS}}\\[6pt]
&+i\epsilon\bigg[\displaystyle{\sum_{n\in\mathbb{N}_0}}\left(
e^{i(n+\frac{1}{2})\phi}\,|n,\,+;\,\eta\rangle\rangle_{\mathrm{R}}+
e^{-i(n+\frac{1}{2})\phi}\,|n,\,-;\,\eta\rangle\rangle_{\mathrm{R}}
\right)\\[4pt]
&\phantom{+i\epsilon\bigg[}+\displaystyle{\sum}_{(a,b)\in
\mathbb{Z}^2\setminus\{(0,0)\}}e^{iAa+iBb}|a,\,b,\,\phi;\,\eta
\rangle\rangle_{\mathrm{R}}\bigg]\bigg\}\,.
\end{array}
\end{equation}
In this notation, $A$ is the relative position of the brane, $B$ its 
Wilson line, and $\epsilon\in\{\pm1\}$ distinguishes a brane from its 
respective anti-brane. The equivalences in this notation are
\begin{eqnarray}
&&\|A+2\pi,\,B,\,\phi,\,\epsilon;\,\eta\rangle\rangle=\|A,\,B+2\pi,\,
\phi,\,\epsilon;\,\eta\rangle\rangle=\|A,\,B,\,\phi,\,\epsilon;\,\eta
\rangle\rangle\,,\nonumber\\
&&\|A,\,B,\,\phi+2\pi,\,\epsilon;\,\eta\rangle\rangle=\|A,\,B,\,\phi,\,
-\epsilon;\,\eta\rangle\rangle\,.\nonumber
\end{eqnarray}
The $\mathbb{Z}_4$ symmetries from above act as
\begin{equation}
\begin{array}{r@{\;=\;}l}
g\|A,\,B,\,\phi,\,\epsilon;\,\eta\rangle\rangle&\|A,\,B,\,\phi+\pi,\,
\epsilon;\,\eta\rangle\rangle\,,\\[5pt]
g'\|A,\,B,\,\phi,\,\epsilon;\,\eta\rangle\rangle&\|-B,\,A,\,\phi+\pi,\,
\epsilon;\,\eta\rangle\rangle\,.
\end{array}
\end{equation}

After these preperations, we now turn to the Gepner description of the 
mirror theory \cite{Lerche:1989cs,Wendland:2005nz}.

\section{$\mathcal{T}^2$ as a Gepner model}\label{sec:Gepner}

The Gepner construction consists of a free conformal field theory 
describing an uncompactified $D$-dimensional space-time, with an 
interior conformal field theory built by means of a tensor product 
of $N=2$ minimal models \cite{Gepner:1987vz}. An orbifold ensures 
that the complete theory is a consistent superstring theory with 
space-time supersymmetry and a modular invariant partition function 
\cite{Gepner:1987qi}.\\

In the following we shall only consider the internal part of the theory, 
namely a tensor product of two minimal models at level $k=2$, which 
together give a central charge $c=3$. In order to relate this theory 
to the torus, we need to perform an orbifold that can be understood in 
terms of a simple current extension 
\cite{Schellekens:1989am,Fuchs:2000fd,Fuchs:2000gv}.

For the tensor product of two minimal models at level $k=2$, there are 
two primary fields that generate the simple current subgroup. In the 
coset notation, these fields are\footnote{see the appendix for our 
conventions on the coset labels.}
\begin{equation}
u=(\Phi^{0,0}_2,\,\Phi^{0,2}_2)\,,\qquad w=(\Phi^{0,2}_0,\,
\Phi^{0,2}_0)\,,
\end{equation}
again for general even $k$.
Projection onto zero monodromy charge with respect to the current $u$ 
amounts to keeping only fields with
\begin{equation}\label{chargeprojection}
m_1+m_2=\frac{k+2}{2}s_1\quad\mathrm{mod}\;k+2\,,
\end{equation}
and the charge projection with respect to the simple current $w$ provides 
the exclusion of NS-R coupling. The simple current extension therefore 
leaves us with the space of states
\begin{equation}\label{spaceofstates}
\displaystyle{\bigoplus_{\substack{[l_1,m_1,s_1],\\ [l_2,m_2,s_2],\\
t\in\mathbb{Z}_{k+2},\\ \tilde{s}_i=s_i\, \mathrm{mod}
\,2}}(l_1,m_1,s_1)\otimes(l_2,m_2,s_2)\otimes(l_1,m_1-2t,\tilde{s}_1)
\otimes(l_2,m_2-2t,\tilde{s}_2)}\,,
\end{equation}
where the sum runs over equivalence classes (denoted by the square brackets) 
of minimal model representations with coset labels $(l,m,s)$, subject to 
fermion alignment $s_1-s_2=0$ mod 2 to prohibit the NS-R coupling, and 
to charge projection (\ref{chargeprojection}) for zero monodromy charge. 
The first two factors in (\ref{spaceofstates}) refer to left-moving and 
the second two to right-moving representations.\\

For $k=2$, the diagonal algebra of a tensor product of two minimal 
models is an $N=2$ algebra at $c=3$, and (\ref{spaceofstates}) 
decomposes into a direct sum of representations of the diagonal 
algebra. The diagonal representations corresponding to highest weight 
vectors of lowest conformal dimension with respect to the construction 
(\ref{spaceofstates}) at $k=2$ can be read off from the low-level expansion 
of the characters.  One finds 
\begin{eqnarray}\label{decomposition}
(\tfrac{1}{4},\,\tfrac{1}{2})\otimes(\tfrac{1}{4},\,\tfrac{1}{2})&=&
(\tfrac{1}{2},\,1)\oplus(1,\,0)\oplus(2,\,0)\oplus\ldots\,,\nonumber\\
(\tfrac{1}{2},\,0)\otimes(0,\,0)&=&(\tfrac{1}{2},\,0)\oplus
(\tfrac{5}{2},\,0)\oplus(\tfrac{5}{2},\,0)\oplus\ldots\,,\\
(\tfrac{1}{8},\,\tfrac{1}{4})\otimes(\tfrac{1}{8}.\,-\tfrac{1}{4})&=&
(\tfrac{1}{4},\,0)\oplus(\tfrac{5}{4},\,0)\oplus(\tfrac{9}{4},\,0)
\oplus(\tfrac{13}{4},\,0)\oplus\ldots\,,\nonumber
\end{eqnarray}
where the left-hand side gives the highest weights and charges of the 
minimal model representations, and the right-hand side those of the 
representations of the diagonal algebra. The other tensor 
products of minimal model representations that appear in the theory 
are linked to those in (\ref{decomposition}) by spectral flow, where 
the same flow parameter is applied on both factors on the left hand side, 
as well as on the representations appearing in the sum on the right-hand 
side. 
From the expansion one can also guess a general formula for the 
decomposition (\ref{decomposition}) (compare with the case considered 
in \cite{Fredenhagen:2006qw}); this is described in the appendix.
However (\ref{decomposition}) already contains all the information we 
are going to need in the following.\\

In the tensor product of minimal models at $k=2$, the primary fields 
with $l=\tfrac{k}{2}=1$ are fixed points under the action of $u^2$. 
Since $u$ generates a cyclic $\mathbb{Z}_4$ subgroup of the simple 
current group, the stabiliser of these fields is isomorphic to 
$\mathbb{Z}_2$. Due to this fixed point we can not directly apply 
the formulae for the tensor product branes from \cite{Brunner:2005fv}, 
but will have to resolve the $S$-matrix \cite{Fuchs:2000fd,Fuchs:2000gv}. 
The formulae for the branes will be given in the following subsections 
(where we will construct the tensor product states in a similar way as 
in \cite{Brunner:2004zd}). 

Before we come to them, let us point out the two $\mathbb{Z}_{k+2}$ 
symmetries in the theory (\ref{spaceofstates}) that we will use later 
on to fix the map between Ishibashi states of the Gepner model and 
Ishibashi states on the torus. The first is the quantum symmetry, i.e. 
the symmetry that is used to undo the orbifold we have just achieved 
by the simple current extension. It divides the space of states into 
the twist sectors $t=0,\ldots,k+1$ by acting as a phase 
$e^{\frac{2\pi i}{k+2}t}$ on the states within the respective sector. 
The second symmetry acts as a phase $e^{\frac{2\pi i}{k+2}m_1}$ on 
a state with left-moving labels 
$(l_1,\,m_1,\,s_1)\otimes(l_2,\,m_2,\,s_2)$. By considering states in 
representations of low-lying ground-state momenta, one can see that 
these two symmetries are in fact just the symmetries $\mathbb{Z}_4$ 
and $\mathbb{Z}_4'$ from section \ref{ssec:torusZ_4}.

Incidentally, the requirement that the quantum $\mathbb{Z}_4$ symmetry
acts geometrically on the torus side requires that we make use of
mirror symmetry and relate the B type branes of the Gepner model 
to A type branes on the torus.

\subsection{Boundary states on the Gepner model}\label{ssec:gepnerbranes}

Supersymmetric B-type boundary states in the theory (\ref{spaceofstates}) 
satisfy the gluing conditions 
\begin{eqnarray}\label{Gepnergluing}
\left(L_n^{(1)}+L_n^{(2)}-\tilde{L}_{-n}^{(1)}-\tilde{L}_{-n}^{(2)}\right)
\|B\rangle\rangle&=&0\,,\nonumber\\
\left(J_n^{(1)}+J_n^{(2)}+\tilde{J}_{-n}^{(1)}+\tilde{J}_{-n}^{(2)}\right)
\|B\rangle\rangle&=&0\,,\\
\left(G_r^{\pm\,(1)}+G_r^{\pm\,(2)}+i\eta (\tilde{G}_{-r}^{\pm\,(1)}+
\tilde{G}_{-r}^{\pm\,(2)})\right)\|B\rangle\rangle&=&0\,.\nonumber
\end{eqnarray}
Among these states we will focus on the tensor product and the 
permutation branes \cite{Brunner:2005fv}, and give a map between them 
and certain boundary states on the torus.

\subsubsection{Tensor product boundary states}\label{ssec:tbranes}

Tensor product branes satisfy (\ref{Gepnergluing}) separately for the 
two tensor product factors $(1)$ and $(2)$. In the sector of the form
\begin{equation}\label{onefactortensorproduct}
(l_i,m_i,s_i)\otimes(l_i,m_i-2t,\tilde{s}_i)\quad(i=1,\,2)
\end{equation}
we find a B-type Ishibashi state if there exists an automorphism between 
the left- and the right-moving representation which takes the coset 
labels $(l,\,m,\,s)$ to
\begin{equation}\label{bautomorphismcondition}
\tilde{l}=l\,,\quad\tilde{m}=-m\,,\quad\tilde{s}=-s
\end{equation}
up to field identification. In other words, we can construct a B-type 
Ishibashi state when a left-moving representation $(l,\,m,\,s)$ is 
tensored to a right-moving representation $(\tilde{l},\,\tilde{m},\,
\tilde{s})$ in (\ref{onefactortensorproduct}), which amounts to 
demanding that
\begin{equation}\label{direct}
\begin{array}{l}
m_i=t\quad \mathrm{mod}\;k+2\,,\\[4pt]
s_i=-\tilde{s}_i\,.
\end{array}
\end{equation}
There is a subtlety when $l_i=\tfrac{k}{2}$, where the labels of the 
right-moving representation in (\ref{onefactortensorproduct}) may encode 
the conjugate representation, but (\ref{bautomorphismcondition}) is only 
met  after a field identification. Note that our convention to use the 
same $l$ labels in the left- and the right-moving representation prevents 
us from overlooking this possibility in the other cases. There exist 
therefore additional Ishibashi states for
\begin{eqnarray}\label{flipped}
l_i&=&\tilde{l}_i=\frac{k}{2}\,,\nonumber\\
m_i&=&t+\frac{k+2}{2}\quad\mathrm{mod}\;k+2\,,\\
s_i&=&-\tilde{s}_i-2\,.\nonumber
\end{eqnarray}
Combination of the `direct' case (\ref{direct}) and the `flipped' case 
(\ref{flipped}) for the two factors $i=1,\,2$ yields the four 
possibilities
\begin{center}
\begin{tabular}{ll}
1. direct-direct&(all values of $l_1,\,l_2$, no field identification 
necessary),\\[2pt]
2. direct-flipped&($l_2=\tfrac{k}{2}$ with field identification, 
all $l_1$),\\[2pt]
3. flipped-direct&($l_1=\tfrac{k}{2}$ with field identification, 
all $l_2$),\\[2pt]
4. flipped-flipped&($l_1=l_2=\tfrac{k}{2}$).
\end{tabular}
\end{center}
Since the automorphism condition is to be matched for both factors 
$i=1,\,2$, the charge projection (\ref{chargeprojection}) gives a 
further restriction on the twist sectors. In the four cases, the charge 
projection is
\begin{enumerate}
\item $2t=\frac{k+2}{2}s_1$ mod $k+2$. From this, we obtain states in the 
Neveu-Schwarz sector ($s_i=0$ mod 2) if $t=0$ mod $\frac{k+2}{2}$; these 
direct states have $m_i=t$, and therefore the $l_i$ will take the values 
$l_i=t$ mod 2. On the other hand, we obtain states in the Ramond sector 
($s_i=1$ mod 2) if $t=\frac{k+2}{4}$ mod $\frac{k+2}{2}$; their $l$ 
labels take the values $l_i=t+1$ mod 2.

\item $2t=\frac{k+2}{2}(s_1+1)$ mod $k+2$. Here the first factor gives 
an Ishibashi state by flipping the right-moving representation, and 
$l_1=\frac{k}{2}$. The $m$ labels are $m_1=t+\frac{k+2}{2}$, and $m_2=t$, 
since the second factor is unflipped. We will get Ramond states for 
$t=0$ mod $\frac{k+2}{2}$, and the alignment of the coset labels of 
the first factor tells us that $t$ has to be even. The label $l_2$ is 
odd. Furthermore, we will get NS states for $t=\frac{k+2}{4}$ mod 
$\frac{k+2}{2}$ odd, where again $l_2$ only takes odd values.

\item the same as in case 2, and we obtain the same result as there 
with interchanged indices ($1\leftrightarrow2$).

\item $2t=\frac{k+2}{2}s_1$ mod $k+2$, which is the same as in case 1, 
but this time both factors have flipped right-moving labels. Therefore, 
$l_i=\frac{k}{2}$, and $m_i=t+\frac{k+2}{2}$ ($i=1,\,2$). This gives 
additional Ishibashi states in the Ramond sector for
$t=\frac{k+2}{4}$ mod $\frac{k+2}{2}$ even, and contributes an additional 
state in the Neveu-Schwarz sector if $t=\frac{k+2}{2}$ is odd.
\end{enumerate}
The labels of representations in which Ishibashi states appear are 
listed in Table \ref{ishibashistatestable} for the different values of $k$.
\begin{table}
\begin{center}
\scalebox{0.99}{\begin{tabular}{|c|ccc|ccc|}
\hline\multicolumn{7}{|c|}{}\\[-11pt]
\multicolumn{7}{|c|}{I. $\frac{k+2}{2}$ odd}\\[4pt]
\hline&&&&&&\\[-11pt]
$t$&$m_1$ mod $k+2$&$s_1$&$l_1$&$m_2$ mod
$k+2$&$s_2$&$l_2$\\[4pt]
\hline&&&&&&\\[-11pt]
0&0&even&even&0&even&even\\[4pt]
&0&odd&odd&$\frac{k+2}{2}$&odd&$\frac{k}{2}$\\[4pt]
&$\frac{k+2}{2}$&odd&$\frac{k}{2}$&0&odd&odd\\[4pt]
\hline&&&&&&\\[-11pt]
$\frac{k+2}{2}$&$\frac{k+2}{2}$&even&odd&$\frac{k+2}{2}$&even&odd \\[4pt]
&0&even&$\frac{k}{2}$&0&even&$\frac{k}{2}$\\[4pt]
\hline

\hline\multicolumn{7}{|c|}{}\\[-11pt]
\multicolumn{7}{|c|}{II. $\frac{k+2}{4}$ odd}\\[4pt]
\hline&&&&&&\\[-11pt]
$t$&$m_1$ mod $k+2$&$s_1$&$l_1$&$m_2$ mod
$k+2$&$s_2$&$l_2$\\[4pt]
\hline&&&&&&\\[-11pt]
0&0&even&even&0&even&even\\[4pt]
&0&odd&odd&$\frac{k+2}{2}$&odd&$\frac{k}{2}$\\[4pt]
&$\frac{k+2}{2}$&odd&$\frac{k}{2}$&0&odd&odd\\[4pt]
\hline&&&&&&\\[-11pt]
$\frac{k+2}{4}$&$\frac{k+2}{4}$&odd&even&$\frac{k+2}{4}$&odd&even\\[4pt]
&$\frac{k+2}{4}$&even&odd&$3\frac{k+2}{4}$&even&$\frac{k}{2}$\\[4pt]
&$3\frac{k+2}{4}$&even&$\frac{k}{2}$&$\frac{k+2}{4}$&even&odd\\[4pt]
\hline&&&&&&\\[-11pt]
$\frac{k+2}{2}$&$\frac{k+2}{2}$&even&even&$\frac{k+2}{2}$&even&even\\[4pt]
&0&odd&$\frac{k}{2}$&$\frac{k+2}{2}$&odd&odd\\[4pt]
&$\frac{k+2}{2}$&odd&odd&0&odd&$\frac{k}{2}$\\[4pt]
\hline&&&&&&\\[-11pt]
$3\frac{k+2}{4}$&$3\frac{k+2}{4}$&odd&even&$3\frac{k+2}{4}$&odd&even\\[4pt]
&$\frac{k+2}{4}$&even&$\frac{k}{2}$&$3\frac{k+2}{4}$&even&odd\\[4pt]
&$3\frac{k+2}{4}$&even&odd&$\frac{k+2}{4}$&even&$\frac{k}{2}$\\[4pt]
\hline

\hline\multicolumn{7}{|c|}{}\\[-11pt]
\multicolumn{7}{|c|}{III. $\frac{k+2}{4}$ even}\\[4pt]
\hline&&&&&&\\[-11pt]
$t$&$m_1$ mod $k+2$&$s_1$&$l_1$&$m_2$ mod
$k+2$&$s_2$&$l_2$\\[4pt]
\hline&&&&&&\\[-11pt]
0&0&even&even&0&even&even\\[4pt]
&0&odd&odd&$\frac{k}{2}$&odd&$\frac{k}{2}$\\[4pt]
&$\frac{k+2}{2}$&odd&$\frac{k}{2}$&0&odd&odd\\[4pt]
\hline&&&&&&\\[-11pt]
$\frac{k+2}{4}$&$\frac{k+2}{4}$&odd&odd&$\frac{k+2}{4}$&odd&odd\\[4pt]
&$3\frac{k+2}{4}$&odd&$\frac{k}{2}$&$3\frac{k+2}{4}$&odd&
$\frac{k}{2}$\\[4pt]
\hline&&&&&&\\[-11pt]
$\frac{k+2}{2}$&$\frac{k+2}{2}$&even&even&$\frac{k+2}{2}$&even&even\\[4pt]
&0&odd&$\frac{k}{2}$&$\frac{k+2}{2}$&odd&odd\\[4pt]
&$\frac{k+2}{2}$&odd&odd&0&odd&$\frac{k}{2}$\\[4pt]
\hline&&&&&&\\[-11pt]
$3\frac{k+2}{4}$&$3\frac{k+2}{4}$&odd&odd&$3\frac{k+2}{4}$&odd&odd\\[4pt]
&$\frac{k+2}{4}$&odd&$\frac{k}{2}$&$\frac{k+2}{4}$&odd&$\frac{k}{2}$\\[4pt]
\hline
\end{tabular}}
\caption{\textit{The possible Ishibashi states for different parities of
$k$.}}\label{ishibashistatestable}
\end{center}
\end{table}
With the $k=2$ model in mind, we will now focus on the case where 
$\frac{k+2}{4}$ is odd, i.e. $k=2$ mod 8.
According to Table \ref{ishibashistatestable}, the Ishibashi states in 
the twist sector labelled by $t=\nu\frac{k+2}{4}$ form three groups,
one where the $l$ labels are both even and the charge labels 
$m_i=\nu\frac{k+2}{4}=t$ show that we are dealing with an `unflipped' 
case in both factors, and two groups where one of the $l$ labels takes 
the value $\frac{k}{2}$ and the corresponding charge label is shifted 
to $m=\frac{k+2}{4}(\nu+2)$, thus indicating a `flipped' case, while 
the other factor has $l$ odd and is unflipped. Note that there are no 
states where it was necessary for both factors to switch the field labels.\\

The standard tensor product branes at $k=2$ mod 8 are given by
\begin{equation}\label{I}
\begin{array}{l}
\|L_1,\,M_1,\,S_1;\,L_2,\,M_2,\,S_2\rangle\rangle=\\
\\
\hspace{2cm}\displaystyle{(k+2)
\sum_{\substack{\nu\in\mathbb{Z}_4,\\s_1,s_2}}}\,
\displaystyle{\sum_{\substack{l_1,l_2\\\mathrm{even}}}}
\frac{S_{L_1,M_1,S_1;l_1,\nu\frac{k+2}{4},s_1}
S_{L_2,M_2,S_2;l_2,\nu\frac{k+2}{4},s_2}}
{\sqrt{S_{0,0,0;l_1,\nu\frac{k+2}{4},s_1}
S_{0,0,0;l_2,\nu\frac{k+2}{4},s_2}}}\\
\\
\multicolumn{1}{r}{\times|l_1,\nu\frac{k+2}{4},s_1;l_2,
\nu\frac{k+2}{4},s_2\rangle\rangle\,,}
\end{array}
\end{equation}
where the $s_i$ obey $l_i+m_i+s_i$ even for $i=1,\,2$. Note that 
these branes only couple to Ishibashi states in representations with 
even $l$ labels; no flipped states are involved so far. The formula 
(\ref{I}) only makes sense for $L_i+M_i+S_i$ even. We will be interested 
in an alignment $\eta_1=\eta_2=\eta$, and hence restrict ourselves to 
$S_1-S_2$ even.\\
There are the following identifications for the brane labels: First, 
we have the analogue of the field identification, 
$(L_i,\,M_i,\,S_i)=(k-L_i,\,M_i+k+2,\,S_i+2)$ for either $i=1$ or 
$i=2$. Second, we have the identification $L_i=k-L_i$, again for either 
$i=1$ or $i=2$. Furthermore, we notice that all branes with the same 
value of $M_1+M_2$ mod 8 are identical, and since $s_1-s_2$ is even, 
we also have $(S_1,\,S_2)=(S_1+2,\,S_2+2)$. Last, a shifting 
$S_1+S_2\mapsto S_1+S_2+2$ is equivalent to shifting 
$M_1+M_2\mapsto M_1+M_2+4$.\\
Given these identifications, we conclude that there are $2k^2$ 
inequivalent branes of type (\ref{I}), $k^2$ for $S_i$ odd and 
$k^2$ for $S_i$ even. In our case, where $k=2$, we will hence have 
8 of these branes, or 4 if we restrict to both $S_1$ and $S_2$ even.\\
The overlap of two of these branes,
\begin{displaymath}
\langle\langle\hat{L}_1,\hat{M}_1,\hat{S}_1;\hat{L}_2,\hat{M}_2,
\hat{S}_2\|q^{\frac{1}{2}(L_0+\tilde{L}_0)-\frac{c}{12}} 
\|L_1,\,M_1,\,S_1;\,L_2,\,M_2,\,S_2\rangle\rangle\,,
\end{displaymath}
can be expressed in the open string sector by means of the modular $S$ 
transformation. The tensor product of representations 
$[l_1',m_1',s_1'] \otimes [l_2',m_2',s_2']$ appears in the open string 
sector with multiplicity
\begin{equation}\label{I-I}
\begin{array}{r@{}l}
\multicolumn{2}{l}{\left(\mathcal{N}_{L_1,l_1'}^{\hat{L}_1}+
\mathcal{N}_{k-L_1,l_1'}^{\hat{L}_1}\right)\left(
\mathcal{N}_{L_2,l_2'}^{\hat{L}_2}+
\mathcal{N}_{k-L_2,l_2'}^{\hat{L}_2}\right)\delta^{(2)}
(S_1-\hat{S}_1+s_1')}\\
&\\
\;\;\;\times\delta^{(2)}(S_2-\hat{S}_2+s_2')\;\delta^{(4)}
\bigg(&\frac{1}{2}(M_1-\hat{M}_1+m_1'+M_2-\hat{M}_2+m_2')\\
&\\
&-(S_1-\hat{S}_1+s_1'+S_2-\hat{S}_2+s_2')\bigg)\,.
\end{array}
\end{equation}
We can see from this formula that the open string vacuum appears with 
multiplicity 2 if either $L_1$ or $L_2$ is equal to $\tfrac{k}{2}$, and 
with multiplicity 4 if both $L_1=L_2=\tfrac{k}{2}$. The branes of the 
first kind, where the open string vacuum is contained twice, must be 
resolved, which yields for $L_1=\tfrac{k}{2}$, $L_2\neq\tfrac{k}{2}$
\begin{equation}\label{IIa}
\begin{array}{l}
\|\frac{k}{2},\,M_1,\,S_1;\,L_2,\,M_2,\,S_2\rangle\rangle=\\
\\
\hspace{2cm}\frac{1}{2}\|\frac{k}{2},\,M_1,\,S_1;\,L_2,\,M_2,\,S_2
\rangle\rangle_{\mathrm{unresolved}}\\
\\
\hspace{2cm}+\frac{k+2}{\sqrt{8}} 
\displaystyle{\sum_{\substack{\nu\in\mathbb{Z}_4,\\s_1,s_2}}}
\displaystyle{\sum_{l\,\mathrm{odd}}}
e^{i\frac{\pi}{4}M_1(\nu+2)-i\frac{\pi}{2}S_1s_1}
\frac{S_{L_2,M_2,S_2;l,\nu\frac{k+2}{4},s_2}}
{\sqrt{S_{0,0,0;l,\nu\frac{k+2}{4},s_2}}}\\
\\
\multicolumn{1}{r}{\times |\frac{k}{2},\frac{k+2}{4}(\nu+2),
s_1;l,\nu\frac{k+2}{4},s_2\rangle\rangle\,,}
\end{array}
\end{equation}
where $\|\frac{k}{2},\,M_1,\,S_1;\,L_2,\,M_2,\,S_2\rangle
\rangle_{\mathrm{unresolved}}$ stands for a boundary state of the 
form (\ref{I}). Note that we can drop the usual factor $\pm 1$ in 
front of the additional part, since this would only give us another 
identification in the set of the boundary state labels, namely 
$(L_2,\pm 1)\equiv(k-L_2,\mp 1)$. The same values of $M_1+M_2$ do not 
necessarily encode the same brane any longer; however, shifting 
$L_2\mapsto k-L_2$ is compensated by $M_1\mapsto M_1+2$. Altogether, 
we find $8k$ different states with $S_1-S_2$ even of this type; $4k$ 
branes with both $S_i$ even, and $4k$ branes with $S_i$ odd. For 
$k=2$, we then have 8 branes with $S_i$ even.\\
For $L_1\neq\tfrac{k}{2}$, $L_2=\tfrac{k}{2}$, we have the analogous 
formula
\begin{equation}\label{IIb}
\begin{array}{l}
\|L_1,\,M_1,\,S_1;\,\frac{k}{2},\,M_2,\,S_2\rangle\rangle=\\
\\
\hspace{2cm}\frac{1}{2}\|L_1,\,M_1,\,S_1;\,\frac{k}{2},\,M_2,\,S_2
\rangle\rangle_{\mathrm{unresolved}}\\
\\
\hspace{2cm}+\frac{k+2}{\sqrt{8}} 
\displaystyle{\sum_{\substack{\nu\in\mathbb{Z}_4,\\s_1,s_2}}}
\displaystyle{\sum_{l\,\mathrm{odd}}}
\frac{S_{L_1,M_1,S_1;l,\nu\frac{k+2}{4},s_1}}
{\sqrt{S_{0,0,0;l,\nu\frac{k+2}{4},s_1}}}
e^{i\frac{\pi}{4}M_2(\nu+2)-i\frac{\pi}{2}S_2s_2}\\
\\
\multicolumn{1}{r}{\times |l,\frac{k+2}{4}\nu,s_1;\frac{k}{2},
\frac{k+2}{4}(\nu+2),s_2\rangle\rangle\,,}
\end{array}
\end{equation}
for again $8k$ different states. The branes (\ref{IIa}), (\ref{IIb}) 
thus get resolved by means of the flipped Ishibashi states; branes with 
$L_1=\tfrac{k}{2},\,L_2\neq\tfrac{k}{2}$ couple to representations where 
we need the field identification in the first factor, and branes with 
$L_1\neq\tfrac{k}{2},\,L_2=\tfrac{k}{2}$ couple to representations that 
are flipped in the second factor. Since there are no states in which we 
had to use the field identification in both factors, it seems reasonable 
that the branes at $L_1=L_2=\tfrac{k}{2}$ do not couple to any flipped 
Ishibashi state. This is indeed the case, and we find the formula
\begin{equation}\label{III}
\|\tfrac{k}{2},M_1,S_1;\tfrac{k}{2},M_2,S_2\rangle\rangle=
\tfrac{1}{2}\,\|\tfrac{k}{2},M_1,S_1;\tfrac{k}{2},M_2,S_2
\rangle\rangle_{\mathrm{unresolved}}
\end{equation}
for these branes. There are 8 different branes of this kind, 4 branes with 
$S_i$ even and 4 with $S_i$ odd.

\subsubsection{Permutation boundary states}\label{ssec:pbranes}

Permutation boundary states of type B satisfy the gluing conditions
\begin{eqnarray}
(L_n^{(1)}-\tilde{L}_{-n}^{(2)})\|B\rangle\rangle=(L_n^{(2)}-
\tilde{L}_{-n}^{(1)})\|B\rangle\rangle&=&0\,,\nonumber\\
(J_n^{(1)}+\tilde{J}_{-n}^{(2)})\|B\rangle\rangle=(J_n^{(2)}+
\tilde{J}_{-n}^{(1)})\|B\rangle\rangle&=&0\,,\nonumber\\
(G_r^{\pm\,(1)}+i\eta \tilde{G}_{-r}^{\pm\,(2)})\|B\rangle\rangle=
(G_r^{\pm\,(2)}+i\eta \tilde{G}_{-r}^{\pm\,(1)})\|B\rangle\rangle&=&
0\,.\nonumber
\end{eqnarray}
Whenever we have to distinguish explicitly between tensor product and 
permutation boundary states we will denote the latter with an 
additional superscript $\sigma$, $\|B\rangle\rangle^{\sigma}$. The 
permutation boundary states have been worked out in 
\cite{Recknagel:2002qq,Gaberdiel:2002jr,Brunner:2005fv}:
\begin{equation}\label{permutationbrane}
\begin{array}{r@{}l}
\displaystyle{\|L,\,M,\,\hat{M},\,S_1,\,S_2\rangle\rangle=
\frac{1}{k+2}\sum_{\nu\in\mathbb{Z}_4}\sum_{l,\,m}
\sum_{s_1,\,s_2}}&\displaystyle{\frac{S_{Ll}}{S_{0l}}
e^{i\frac{\pi}{4}\hat{M}\nu+i\frac{\pi}{k+2}Mm-i\frac{\pi}{2} 
(S_1s_1+S_2s_2)}}\\[4pt]
&\displaystyle{\times |l,\,m+n,\,s_1;\,l,\,-m+n,\,s_2\rangle\rangle\,,}
\end{array}
\end{equation}
where the sums over $l$ and $m$ run over appropriate values in the twist 
sector $n=\nu\tfrac{k+2}{4}$, and $s_i=l+m+n$ mod 2. In this formula, we 
have the label constraints $L+M+S_1+S_2=0$ mod 2 and $M-\hat{M}=0$ mod 2, 
and we restrict ourselves again to states with $S_1-S_2=0$ mod 2.\\

For $L\neq\tfrac{k}{2}$, there is again an analogue of the field 
identification, $(L,\,M,\,\hat{M},\,S_1+S_2)=(k-L,\,M+k+2,\,\hat{M}+4,\,
S_1+S_2+2)$. From the charge projection (\ref{chargeprojection}) we see 
that $(\hat{M},\,S_1+S_2)=(\hat{M}+4,\,S_1+S_2+2)$, which we can combine 
with the field identification to yield $(L,\,M)=(k-L,\,M+k+2)$. 
Furthermore, states with $(S_1,\,S_2)$ and $(S_1+2,\,S_2+2)$ are again 
the same. We conclude that there are $4k(k+2)$ different permutation 
branes with $L\neq\tfrac{k}{2}$ and $S_1-S_2$ even. If $k=2$ there are 
thus 16 different branes with $L\neq 1$ and even $S_i$.

For $L=\tfrac{k}{2}$, $(L,\,M)=(k-L,\,M+k+2)$ is an identification on 
its own, without making use of $(\hat{M},\,S_1+S_2)\mapsto 
(\hat{M}+4,\,S_1+S_2+2)$. There are hence $4(k+2)$ different permutation 
branes with $L=\tfrac{k}{2}$ and $S_1-S_2$ even; for $k=2$ this leaves 
us with 8 different $L=1$ branes at even $S_i$.

Altogether, there are $4k^2+12k+8$ different permutation branes with 
$S_1-S_2$ even for $k=2$ mod 8, compared to a total of $2k^2+16k+8$ 
tensor product branes.

\section{Comparison of torus and Gepner model}\label{sec:comparison}

We will now focus on the case where $k=2$, and compare the boundary 
states we have just described with those of the torus from section 
\ref{ssec:torusbranes}.

In order to compare the two respective classes of Ishibashi states we 
will write the Gepner model Ishibashi states in terms of Ishibashi 
states of the diagonal $N=2$ algebra. An Ishibashi state in the 
left-moving representation
\begin{equation}\label{diagonalrepresentations}
(h_1,\,q_1)\otimes(h_2,\,q_2)=\bigoplus_{[(H,\,Q)]}(H,\,Q)\,,
\end{equation}
where the direct sum runs over the diagonal representations of highest 
weight $H$ and charge $Q$ (see (\ref{decomposition})), consists of a 
sum of `diagonal' Ishibashi states up to the choice of phases 
$\psi_{(H,Q)}$,
\begin{equation}\label{generalishibashistate}
|h_1,\,q_1,\,h_2,\,q_2\rangle\rangle=\sum_{[(H,Q)]}e^{i\psi_{(H,Q)}}
|H,\,Q\rangle\rangle_{(h_1,q_1)\otimes(h_2,q_2)}\,.
\end{equation}
In general we can not always set these phases to zero, since there exist 
tensor products of minimal model representations that admit both a 
tensor product and a permutation Ishibashi state, and for those the 
phases $e^{i\psi_{(H,Q)}}$ have to be different. This is the case 
whenever $h_1=h_2$ and $q_1=q_2$ in (\ref{diagonalrepresentations}).
Let us define the diagonal Ishibashi states in 
(\ref{generalishibashistate}) for $h_1=h_2$ and $q_1=q_2$ such that 
all phases $\psi_{(H,Q)}$ vanish for the permutation Ishibashi state. 
Then, as explained in \cite{Brunner:2005fv}, the phase 
$\psi_{(2h_1,2q_1)}$ of the ground state in the tensor product 
Ishibashi state is
\begin{equation}\label{permutationtensorphase3}
\psi_{(2h_1,2q_1)} = \frac{\pi}{2}s-\frac{\pi}{k+2}m\,,
\end{equation}
where $(l,\,m,\,s)$ are the coset labels of the representation 
$(h_1,\,q_1)$.

\subsection{Brane dictionary from the $\mathbb{Z}_4$ symmetries}
\label{ssec:Z_4}

The identification of the Ishibashi states maps diagonal $N=2$ Ishibashi 
states of the Gepner model to $N=2$ Ishibashi states of the torus with 
the same highest weight and charge. In the vacuum sector, this is already 
sufficient to identify the Ishibashi states. Consider for example the 
diagonal Ishibashi state at $H=\tfrac{1}{2},\,Q=1$, which appears on the 
Gepner side only in the left-moving representation 
$h_1=h_2=\tfrac{1}{4},\,q_1=q_2=\tfrac{1}{2}$ in the twist sector $t=2$. 
Since this is the only Ishibashi state with these quantum numbers, we can 
identify it with the state $|\tfrac{1}{2},\,1\rangle\rangle$ on the 
torus\footnote{Strictly speaking this only fixes the identification up to 
a phase. As we shall see, it is consistent that we choose this phase factor 
to be trivial.}. On the Gepner side, this Ishibashi state couples to a 
permutation brane $(L,\,M,\,\hat{M},\,S_1,\,S_2)$ with the factor
\begin{displaymath}
\frac{1}{\sqrt{2}}\sin\left(\tfrac{\pi}{4}(L+1)\right)
e^{i\frac{\pi}{2}\hat{M}}\,,
\end{displaymath}
while on the torus side the coupling is $\mathcal{N}(\phi)e^{i\pi\phi}$. 
Hence we deduce that the angle of the permutation brane is
\begin{equation}\label{permutationangle}
\phi=\frac{\pi}{2}\hat{M}\,.
\end{equation}

On the Gepner side, we also find a tensor product Ishibashi state in 
the same representation, whose coefficient will analogously yield the 
angle $\phi$ in terms of the tensor product brane labels. Remembering 
the additional phase (\ref{permutationtensorphase3}), we find with the 
coefficients from the formulae (\ref{I}) -- (\ref{III}) 
\begin{equation}
\phi=\frac{\pi}{2}(M_1+M_2+1)
\end{equation}
for a tensor product brane $(L_1,\,M_1,\,S_1,\,L_2,\,M_2,\,S_2)$.

In the more general cases of vanishing diagonal charge ($Q=0$), the 
identification of the diagonal Gepner Ishibashi states at highest 
weight $H>0$ with the torus Ishibashi states is more complicated. 
However we can use that eigenstates of the two $\mathbb{Z}_4$ symmetries 
on the Gepner side will be mapped to eigenstates of the corresponding 
symmetries on the torus side.

As an example, consider the identification of the diagonal Ishibashi 
states of lowest nonvanishing highest weight in the NS sector, which 
are the states $H=\tfrac{1}{4},\,Q=0$. In the Gepner model, these states 
appear in the left-moving representations $(\tfrac{1}{8},\,
\pm\tfrac{1}{4})\otimes (\tfrac{1}{8},\,\mp\tfrac{1}{4})$, whose Ishibashi 
states are of permutation type in the twist sectors $n=0$ and $n=2$ and of 
tensor product type in the other sectors. Let us choose the basis on the 
torus to be
\begin{equation}
|1,\,0,\,\pi;\,\eta\rangle\rangle^{t,1},\qquad |1,\,0,\,\pi;\,
\eta\rangle\rangle^{t,3}\qquad(t=0,\,\ldots,3)
\end{equation}
in the notation (\ref{allZ_4eigenstate}). The ansatz for the identification 
is then 
\begin{eqnarray}\label{ansatz}
&&|(\tfrac{1}{8},\,-\tfrac{1}{4})\otimes(\tfrac{1}{8},\,\tfrac{1}{4})
\rangle\rangle^t\quad\leftrightarrow\quad\alpha^{(t)}|1,\,0,\,\pi;\,
\eta\rangle\rangle^{t,1}\,,\nonumber\\
\\[-11pt] &&|(\tfrac{1}{8},\,\tfrac{1}{4})\otimes(\tfrac{1}{8},\,-
\tfrac{1}{4})\rangle\rangle^t\quad\leftrightarrow\quad\beta^{(t)}
|1,\,0,\,\pi;\,\eta\rangle\rangle^{t,3}\nonumber
\end{eqnarray}
for $0\leq t\leq 3$, where the phases $\alpha^{(t)}$ and $\beta^{(t)}$ 
are initially undetermined. For a permutation brane 
$(L,\,M,\,\hat{M},\,S_1,\,S_2)$ we then obtain
\begin{eqnarray}\label{consistencyequations}
&&e^{iA}=e^{-iA}=\tfrac{1}{\sqrt{2}}e^{i\frac{\pi}{4}M+i\frac{\pi}{2}
L-i\pi S_2}\big(\alpha^{(0)}+\beta^{(0)}e^{i\frac{\pi}{2}M}\big)\,,
\nonumber\\
&&e^{iB}=e^{-iB}=\tfrac{i}{\sqrt{2}}e^{i\frac{\pi}{4}M+i\frac{\pi}{2}L-
i\pi S_2}\big(\alpha^{(0)}-\beta^{(0)}e^{i\frac{\pi}{2}M}\big)\,,\\
&&\alpha^{(2)}=-\beta^{(0)}\,,\quad \beta^{(2)}=-\alpha^{(0)}\,.\nonumber
\end{eqnarray}
Here, $A$ and $B$ are position and Wilson line of the torus brane. The label 
$L$ is even, since these are the only permutation branes that couple to the 
considered Ishibashi states. A solution to the equations 
(\ref{consistencyequations}), i.e. a consistent formula for position $A$ and 
Wilson line $B$ in terms of the permutation brane labels, can be given for 
$\alpha^{(0)}=(\beta^{(0)})^*=-(\alpha^{(2)})^*=-\beta^{(2)}=e^{i\frac{\pi}{4}}$ 
(see Table \ref{identificationtable})\footnote{Again, there exist other 
possible phase choices. These correspond to choosing the absolute position
and orientation of one reference brane.}. We find similar consistency 
equations for $A$  and $B$ from the tensor product branes at a single fixed 
point (\ref{IIa}), (\ref{IIb}).

Positions and Wilson lines of the other tensor product and permutation branes 
(at $L=1$ or $L_1+L_2$ even, respectively) can be obtained from matching the 
states at $H=\tfrac{1}{2}$, $Q=0$ in a similar way as in the case 
$H=\tfrac{1}{4}$, $Q=0$ we have just mentioned. The formulae for positions 
and Wilson lines were also checked in the R sector, and for higher values 
of $H$.\\

\begin{table}
\begin{center}
\begin{tabular}{|c|c|}
\hline\multicolumn{2}{|l|}{}\\[-11pt]
\multicolumn{2}{|l|}{Permutation branes}\\[3pt]
\hline\multicolumn{2}{|l|}{}\\[-11pt]
\multicolumn{2}{|c|}{$\phi=\tfrac{\pi}{2}\hat{M}$}\\[3pt]
\hline&\\[-11pt]
$L=0$ mod 2 & $\begin{array}{ccc}
M=0\;\mathrm{mod}\;4\,: & A = \tfrac{\pi}{2}L+\tfrac{\pi}{4}M+\pi S_1\,,&
B=A\\
M=2\;\mathrm{mod}\;4\,: & A = \tfrac{\pi}{2}L+\tfrac{\pi}{4}(M-2)+
\pi S_1\,,&B=A+\pi
\end{array}$\\[3pt]
\hline&\\[-11pt]
$L=1$ & $A=\tfrac{\pi}{2}(M-1)\,,\;B=A+\pi$\\[3pt]
\hline
\multicolumn{2}{c}{}\\
\hline\multicolumn{2}{|l|}{}\\[-11pt]
\multicolumn{2}{|l|}{Tensor product branes}\\[3pt]
\hline\multicolumn{2}{|l|}{}\\[-11pt]
\multicolumn{2}{|c|}{$\phi=\tfrac{\pi}{2}(M_1+M_2+1)$}\\[3pt]
\hline&\\[-11pt]
$L_1=1$, $L_2=0$ mod 2 & $A=\tfrac{\pi}{2}L_2+\tfrac{\pi}{2}M_1+\pi S_1
\,,\;B=A$\\[3pt]
\hline&\\[-11pt]
$L_1=0$ mod 2, $L_2=1$ & $A=\tfrac{\pi}{2}L_1+\tfrac{\pi}{2}M_2+\pi 
(S_1+1)\,,\;B=A+\pi$\\[3pt]
\hline&\\[-11pt]
$L_1,\,L_2=0$ mod 2 & $A = B = 0$\\[3pt]
\hline&\\[-11pt]
$L_1=L_2=1$ & $A=B=\pi$\\[3pt]
\hline
\end{tabular}
\end{center}
\caption{\textit{Example for a consistent choice of positions $A$ and 
Wilson lines $B$ in terms of the coset labels for the images of the 
B-type permutation and tensor product branes, with 
$\epsilon=e^{-i\frac{\pi}{2} (S_1+S_2)}$.}}\label{identificationtable}
\end{table}
\begin{table}
\begin{pspicture}(-1,-1)(4.5,4.5)
\psline[linewidth=0.8pt,linestyle=dotted,linecolor=gray](0,1)(4,1)
\psline[linewidth=0.8pt,linestyle=dotted,linecolor=gray](0,2)(4,2)
\psline[linewidth=0.8pt,linestyle=dotted,linecolor=gray](0,3)(4,3)
\psline[linewidth=0.8pt,linestyle=dotted,linecolor=gray](0,4)(4,4)
\psline[linewidth=0.8pt,linestyle=dotted,linecolor=gray](1,0)(1,4)
\psline[linewidth=0.8pt,linestyle=dotted,linecolor=gray](2,0)(2,4)
\psline[linewidth=0.8pt,linestyle=dotted,linecolor=gray](3,0)(3,4)
\psline[linewidth=0.8pt,linestyle=dotted,linecolor=gray](4,0)(4,4)
\psline[linewidth=1.5pt,linecolor=gray]{->}(0,0)(0,4.2)
\psline[linewidth=1.5pt,linecolor=gray]{->}(0,0)(4.2,0)
\gray{%
\rput(2,-0.5){$\pi$}
\rput(4,-0.5){$2\pi$}
\rput(-0.5,2){$\pi$}
\rput(-0.5,4){$2\pi$}}
\rput(3.2,0.3){\psframebox*{\scalebox{0.7}{\black $M=0\,,\hat{M}=0$}}}
\psline[linewidth=3pt](0,0)(4,0)
\rput(3.2,2.3){\psframebox*{\scalebox{0.7}{\black $M=2\,,\hat{M}=0$}}}
\psline[linewidth=3pt](0,2)(4,2)
\rput{90}(0.3,3.1){\psframebox*{\scalebox{0.7}{\black $M=0\,,\hat{M}=2$}}}
\psline[linewidth=3pt](0,0)(0,4)
\rput{90}(1.7,3.1){\psframebox*{\scalebox{0.7}{\black $M=2\,,\hat{M}=2$}}}
\psline[linewidth=3pt](2,0)(2,4)
\pscircle[fillstyle=solid](4,0){0.15}
\pscircle[fillstyle=solid](2,4){0.15}
\pscircle[fillstyle=solid](4,2){0.15}
\pscircle[fillstyle=solid](0,4){0.15}
\end{pspicture}
\hspace{1cm}
\begin{pspicture}(-1,-1)(4.5,4.5)
\psline[linewidth=0.8pt,linestyle=dotted,linecolor=gray](0,1)(4,1)
\psline[linewidth=0.8pt,linestyle=dotted,linecolor=gray](0,2)(4,2)
\psline[linewidth=0.8pt,linestyle=dotted,linecolor=gray](0,3)(4,3)
\psline[linewidth=0.8pt,linestyle=dotted,linecolor=gray](0,4)(4,4)
\psline[linewidth=0.8pt,linestyle=dotted,linecolor=gray](1,0)(1,4)
\psline[linewidth=0.8pt,linestyle=dotted,linecolor=gray](2,0)(2,4)
\psline[linewidth=0.8pt,linestyle=dotted,linecolor=gray](3,0)(3,4)
\psline[linewidth=0.8pt,linestyle=dotted,linecolor=gray](4,0)(4,4)
\psline[linewidth=1.5pt,linecolor=gray]{->}(0,0)(0,4.2)
\psline[linewidth=1.5pt,linecolor=gray]{->}(0,0)(4.2,0)
\gray{%
\rput(2,-0.5){$\pi$}
\rput(4,-0.5){$2\pi$}
\rput(-0.5,2){$\pi$}
\rput(-0.5,4){$2\pi$}}
\rput{45}(2.8,3.1){\psframebox*{\scalebox{0.7}{\black $M=1,\,\hat{M}=1$}}}
\psline[linewidth=3pt](0,0)(4,4)
\rput{-45}(3,1.4){\psframebox*{\scalebox{0.7}{\black $M=1,\,\hat{M}=3$}}}
\psline[linewidth=3pt](0,4)(4,0)
\pscircle[fillstyle=solid](4,0){0.15}
\pscircle[fillstyle=solid](4,4){0.15}
\pswedge[fillstyle=solid,fillcolor=black](4,0){0.15}{90}{-90}
\pswedge[fillstyle=solid,fillcolor=black](4,4){0.15}{90}{-90}
\end{pspicture}
\caption{\textit{Permutation brane positions in the labels of Table }
\ref{identificationtable}.\textit{ The left diagram shows the short 
branes $(L=0)$ at different values of $M$ and $\hat{M}$, the right 
diagram contains examples of the long branes $(L=1)$. 
The fermion structure has been set to $\eta=+1$, $S_1=S_2=0$. The 
filling of the circles at the end of the lines denotes the Wilson line 
of the brane; empty circles correspond to Wilson line $B=0$, half-filled 
circles to Wilson line $B=\pi$.}}\label{Permutationbranetable}
\end{table}

\begin{table}
\begin{pspicture}(-1,-1)(4.5,4.5)
\psline[linewidth=0.8pt,linestyle=dotted,linecolor=gray](0,1)(4,1)
\psline[linewidth=0.8pt,linestyle=dotted,linecolor=gray](0,2)(4,2)
\psline[linewidth=0.8pt,linestyle=dotted,linecolor=gray](0,3)(4,3)
\psline[linewidth=0.8pt,linestyle=dotted,linecolor=gray](0,4)(4,4)
\psline[linewidth=0.8pt,linestyle=dotted,linecolor=gray](1,0)(1,4)
\psline[linewidth=0.8pt,linestyle=dotted,linecolor=gray](2,0)(2,4)
\psline[linewidth=0.8pt,linestyle=dotted,linecolor=gray](3,0)(3,4)
\psline[linewidth=0.8pt,linestyle=dotted,linecolor=gray](4,0)(4,4)
\psline[linewidth=1.5pt,linecolor=gray]{->}(0,0)(0,4.2)
\psline[linewidth=1.5pt,linecolor=gray]{->}(0,0)(4.2,0)
\gray{%
\rput(2,-0.5){$\pi$}
\rput(4,-0.5){$2\pi$}
\rput(-0.5,2){$\pi$}
\rput(-0.5,4){$2\pi$}}
\rput(2.0,0.7){\psframebox*{\scalebox{0.6}{\black $(M_1,M_2)=(3,0)$}}}
\rput(2.0,3.3){\psframebox*{\scalebox{0.6}{\black $(M_1,M_2)=(1,2)$}}}
\rput{90}(0.7,2.0){\psframebox*{\scalebox{0.6}{\black $(M_1,M_2)=(1,0)$}}}
\rput{90}(3.3,2.0){\psframebox*{\scalebox{0.6}{\black $(M_1,M_2)=(3,2)$}}}
\psline[linewidth=3pt](3,0)(3,4)
\psline[linewidth=3pt](0,1)(4,1)
\psline[linewidth=3pt](0,3)(4,3)
\psline[linewidth=3pt](1,0)(1,4)
\pscircle[fillstyle=solid](4,1){0.15}
\pscircle[fillstyle=solid](4,3){0.15}
\pscircle[fillstyle=solid](1,4){0.15}
\pscircle[fillstyle=solid](3,4){0.15}
\pswedge[fillstyle=solid,fillcolor=black](4,1){0.15}{90}{180}
\pswedge[fillstyle=solid,fillcolor=black](4,3){0.15}{90}{0}
\pswedge[fillstyle=solid,fillcolor=black](1,4){0.15}{90}{180}
\pswedge[fillstyle=solid,fillcolor=black](3,4){0.15}{90}{0}
\end{pspicture}
\hspace{1cm}
\begin{pspicture}(-1,-1)(4.5,4.5)
\psline[linewidth=0.8pt,linestyle=dotted,linecolor=gray](0,1)(4,1)
\psline[linewidth=0.8pt,linestyle=dotted,linecolor=gray](0,2)(4,2)
\psline[linewidth=0.8pt,linestyle=dotted,linecolor=gray](0,3)(4,3)
\psline[linewidth=0.8pt,linestyle=dotted,linecolor=gray](0,4)(4,4)
\psline[linewidth=0.8pt,linestyle=dotted,linecolor=gray](1,0)(1,4)
\psline[linewidth=0.8pt,linestyle=dotted,linecolor=gray](2,0)(2,4)
\psline[linewidth=0.8pt,linestyle=dotted,linecolor=gray](3,0)(3,4)
\psline[linewidth=0.8pt,linestyle=dotted,linecolor=gray](4,0)(4,4)
\psline[linewidth=1.5pt,linecolor=gray]{->}(0,0)(0,4.2)
\psline[linewidth=1.5pt,linecolor=gray]{->}(0,0)(4.2,0)
\gray{%
\rput(2,-0.5){$\pi$}
\rput(4,-0.5){$2\pi$}
\rput(-0.5,2){$\pi$}
\rput(-0.5,4){$2\pi$}}
\rput{45}(2.8,3.1){\psframebox*{\scalebox{0.7}{\black $M_1=0,\,M_2=0$}}}
\psline[linewidth=3pt](0,0)(4,4)
\rput{-45}(3,1.4){\psframebox*{\scalebox{0.7}{\black $M_1=2,\,M_2=2$}}}
\psline[linewidth=3pt](0,4)(4,0)
\pscircle[fillstyle=solid](4,0){0.15}
\pscircle[fillstyle=solid](4,4){0.15}
\end{pspicture}
\caption{\textit{Tensor product brane positions in the labels of Table }
\ref{identificationtable}.\textit{ The left diagram shows the resolved 
short tensor product branes with $L_1=1,\,L_2=0$, the right diagram 
contains long branes with $L_1=L_2=0$. The fermion structure has been 
set to $\eta=+1$, $S_1=S_2=0$. The filling of the circles at the end 
of the lines denotes the Wilson line of the brane; empty circles 
correspond to Wilson line $B=0$, quarter-filled circles to Wilson line 
$B=\tfrac{\pi}{2}$, etc.}}\label{Tensorproductbranetable}
\end{table}

The procedure provides a consistent map for the positions and Wilson 
lines of the images of tensor product and permutation branes on the 
torus, which -- for a certain phase chioce -- is given in Table \ref{identificationtable}. From this 
table, we see that the branes coupling to the Ishibashi states at 
lowest momentum (highest weight $H=\tfrac{1}{4}$) have $L$ even 
(permutation branes) or $L_1+L_2$ odd (resolved tensor product branes 
coupling to flipped states). These branes are the `short' or `light' 
branes, i.e. those that couple to the vacuum with the lowest 
coefficient; their angles are integer multiples of $\pi$. For the 
permutation branes at $L$ even, shifting the $M$ label by 2 leads 
to a relative phase shift between position and Wilson line. In the 
case of the tensor product branes at $L_1+L_2$ odd, the relative 
phase between position and Wilson line is changed by passing from 
the branes with $L_1=1$ to those with $L_2=1$ (and vice versa).

\section{Relation to matrix factorisations}\label{sec:mf}

Our simple model corresponds to an orbifold of the Landau-Ginzburg 
superpotential 
\begin{equation}
W=x_1^4+x_2^4+z^2\,,
\end{equation}
where the presence of the trivial factor $z^2$ is related to the charge 
projection of our Gepner model. Topological branes are given by a pair
\begin{equation}
\left(Q=\left(\begin{array}{cc}0&J\\E&0\end{array}\right),\; \gamma\right)
\end{equation}
where $Q$ has entries that are polynomials in $x_1$, $x_2$, and $z$ such 
that it factorises the superpotential, i.e. 
\begin{equation}
Q^2=W\mathbf{1}\,.
\end{equation}
The orbifold matrix $\gamma$ satisfies
\begin{equation}
\gamma Q(ix_1,ix_2,-z)\gamma^{-1}=Q(x_1,x_2,z)\qquad\mathrm{and}\qquad
\gamma^4=\mathbf{1}\,.
\end{equation}
The orbifold matrix is hence only defined up to a phase factor 
$e^{i\frac{\pi}{2}n}$ for $n=0,\,1,\,2,\,3$. 

The relative couplings of factorisations to the RR-primary ground states 
can be computed from a general formula given in \cite{Kapustin:2003ga} 
(for a review, see e.g. \cite{Walcher:2004tx}). In our case, these states 
are in the following left-moving representations:
\begin{equation}
\begin{array}{ccc}
t & (l_1,\,m_1,\,s_1)\otimes(l_2,\,m_2,\,s_2) & (h_1,\,q_1)\otimes
(h_2,\,q_2)\\[5pt]
1 & (0,\,1,\,1)\otimes(0,\,1,\,1) & (\frac{1}{16},\frac{1}{4})\otimes
(\frac{1}{16},\frac{1}{4})\\
3 & (0,\,7,\,3)\otimes(0,\,7,\,3) & (\frac{1}{16},-\frac{1}{4})\otimes
(\frac{1}{16},-\frac{1}{4})
\end{array}
\end{equation}
As above, $t$ denotes the twist sector. The R primary field with 
$(h,\,q)=(\tfrac{1}{16},\,0)$ appears only in combination with the field 
$(h,\,q)=(\tfrac{5}{16},\,\tfrac{1}{2})$, which is not primary.
Since the RR primary states all appear in twisted sectors, the formula for 
their brane couplings reduces to
\begin{equation}
C(Q,\,\gamma;\,t)=\mathrm{Str}(\gamma^t)\qquad (t =1,\,3)\,,
\end{equation}
Str denoting the supertrace. In the following we are going to identify 
the Gepner branes of section \ref{ssec:gepnerbranes} with certain matrix 
factorisations by computing the couplings of different factorisations 
and comparing the results to the couplings obtained from the brane 
formulae. From now on we will set $S_1=S_2=0$.\\

Let us first consider the matrix factorisations corresponding to the 
permutation branes (\ref{permutationbrane}), which have been worked out 
in general in \cite{Brunner:2005fv}. The analogues of the rank 1 
factorisations described in \cite{Brunner:2005fv} are given by matrices 
of the form
\begin{eqnarray}\label{rank1}
J&=&\left(\begin{array}{cc}\displaystyle{\prod_{\eta\in\mathcal{I}}}
(x_1-\eta x_2)&-z\\z&\displaystyle{\prod_{\eta\in\mathcal{I}^C}}
(x_1-\eta x_2)\end{array}\right)\,,\nonumber\\ E&=&\left(\begin{array}{cc}
\displaystyle{\prod_{\eta\in\mathcal{I}^C}}(x_1-\eta x_2)&z\\
-z&\displaystyle{\prod_{\eta\in\mathcal{I}}}(x_1-\eta x_2)
\end{array}\right)\,,\\
\gamma&=&\mathrm{diag}(1,-i^{|\mathcal{I}|},i^{|\mathcal{I}|},-1)
\times e^{i\frac{\pi}{2}n}\,,\nonumber
\end{eqnarray}
where $\mathcal{I}$ is a subset of the set of fourth roots of 
$-1$, $\mathcal{I}^C$ is its respective complement, and $|\mathcal{I}|$ 
is the number of elements in $\mathcal{I}$. These factorisations are 
identified with branes in the Gepner model in the following way:

\begin{list}{(\roman{enumi})}{\usecounter{enumi}}
\item Factorisations of type (\ref{rank1}) with $|\mathcal{I}|=1$ or 
$|\mathcal{I}|=3$ correspond to the permutation branes 
(\ref{permutationbrane}) with $L\neq 1$. There are 8 factorisations of 
this type, and each has 4 values for the phase of $\gamma$. A 
factorisation with $|\mathcal{I}|=3$ and phase $e^{i\frac{\pi}{2}n}$ 
is identical to a factorisation with $\mathcal{I}^C$ and phase 
$e^{i\frac{\pi}{2}(n+1)}$, so that we are left with 16 different branes, 
as we have expected from the counting in \ref{ssec:pbranes}. Without 
loss of generality, we can restrict ourselves to permutation branes with 
$L=0$. Taking e.g. $\|L=0,\,M=0,\,\hat{M}=0,\,S_i=0\rangle\rangle$ to 
be the factorisation with $\mathcal{I}=\{e^{i\frac{\pi}{4}}\}$ and $n=0$, 
we find that the $L=0$ branes correspond to factorisations 
$|\mathcal{I}|=1$ with $\hat{M}=2n$. The values of $M\in\{0,\,2,\,4,\,6\}$ 
correspond to the choice of the element in $\mathcal{I}$.

\item The factorisations of type (\ref{rank1}), where $\mathcal{I}$ contains 
two consecutive roots of $-1$, correspond to the permutation branes 
(\ref{permutationbrane}) with $L=1$. There are two of these factorisations, 
each with four choices of $\gamma$, and they are again pairwise identified 
in a similar way as before, so that there are 4 different branes. An 
identification consistent with the one of the $L=0$ permutation branes 
from above yields $\hat{M}=2n+1$.

\item The missing factorisations of type (\ref{rank1}) correspond to the 
resolved tensor product branes at $L_1=L_2=1$, as it has already been argued 
on general grounds in \cite{Caviezel:2005th}. The missing factorisations are 
those two for which $\mathcal{I}$ contains two non-consecutive roots, and 
each factorisation has 4 possibilities for $\gamma$. As before there is 
again an identification between pairs that reduces the number of different 
branes to 4, and the phase $e^{i\frac{\pi}{2}n}$ of $\gamma$ is linked to 
the labels $M_i$ by $M_1+M_2=2n$ in our conventions.
\end{list}
We have hence found corresponding Gepner branes for all factorisations 
(\ref{rank1}), in agreement with the proposition in 
\cite{Caviezel:2005th}.\\

Let us also give the factorisations corresponding to the other tensor 
product branes. The branes (\ref{I}) with $L_1=L_2=0$ belong to the 
usual rank 4 tensor product factorisations $x_1x_1^3+x_2x_2^3+zz$. 
There are four factorisations of this type, and each comes again with 
four choices of $\gamma$. There are however only four inequivalent 
factorisations, which are given by
\begin{eqnarray}
&J=\left(\begin{array}{cccc}x_1&-x_2^3&-z&0\\x_2&x_1^3&0&-z\\
z&0&x_1^3&x_2^3\\0&z&-x_2&x_1\end{array}\right)\,,\qquad
E=\left(\begin{array}{cccc}x_1^3&x_2^3&z&0\\-x_2&x_1&0&z\\
-z&0&x_1&-x_2^3\\0&-z&x_2&x_1^3\end{array}\right)\,,&\nonumber\\
\\[-12pt]
&\gamma=\mathrm{diag}(1,1,-i,i,i,-i,-1,-1)\times e^{i\frac{\pi}{2}n}
\,.&\nonumber
\end{eqnarray}
In our conventions we then have $M_1+M_2=2n$.\\

The resolved tensor product branes (\ref{IIa}) with $L_1=1,\,L_2\neq 1$ 
correspond to the rank 2 factorisations $(x_1^2+iz)(x_1^2-iz)+x_2x_2^3$. 
There are four factorisations of this type, each with four choices of 
$\gamma$. They are given by
\begin{eqnarray}\label{MFresolvedI}
&J=\left(\begin{array}{cc}x_1^2-iz&-x_2^3\\x_2&x_1^2+iz 
\end{array}\right)\,,\qquad
E=\left(\begin{array}{cc}x_1^2+iz&x_2^3\\-x_2&x_1^2-iz 
\end{array}\right)\,,&\nonumber\\
\\[-12pt]
&\gamma=\mathrm{diag}(1,i,-1,-i)\times e^{i\frac{\pi}{2}n}
\,,&\nonumber
\end{eqnarray}
and
\begin{eqnarray}\label{MFresolvedII}
&J=\left(\begin{array}{cc}x_1^2+iz&-x_2^3\\x_2&x_1^2-iz \end{array}\right)\,,
\qquad
E=\left(\begin{array}{cc}x_1^2-iz&x_2^3\\-x_2&x_1^2+iz \end{array}\right)\,,&
\nonumber\\
\\[-12pt]
&\gamma=\mathrm{diag}(1,i,-1,-i)\times e^{i\frac{\pi}{2}n}\,,&\nonumber
\end{eqnarray}
with the two other factorisations arising from (\ref{MFresolvedI}), 
(\ref{MFresolvedII}) by interchanging $x_2\leftrightarrow x_2^3$ and 
$(x_1^2+iz)\leftrightarrow (x_1^2-iz)$. However, this interchange leads 
to equivalent factorisations. We hence find 8 different factorisations, 
in agreement with the 8 different resolved tensor product branes with 
$L_1=1,\,L_2\neq1$. In our conventions, we have $M_1+M_2=2n+1$.

The other class (\ref{IIb}) of resolved tensor product branes with 
$L_1\neq 1,\,L_2=1$ can be identified with the factorisations 
$x_1x_1^3+(x_2^2+iz)(x_2^2-iz)$, where the different branes are given 
by (\ref{MFresolvedI}) and (\ref{MFresolvedII}) with 
$x_1\leftrightarrow x_2$.\\

We have thus identified matrix factorisations for all the Gepner 
branes described in \ref{ssec:gepnerbranes}.

\section{Conclusion}\label{sec:conclusion}

In this paper, we have worked out a dictionary between explicit sets 
of tensor product and permutation branes (\ref{I}), (\ref{IIa}), 
(\ref{IIb}), (\ref{III}), (\ref{permutationbrane}) in the Gepner 
construction involving two minimal models at $k=2$ (\ref{spaceofstates}), 
and the branes (\ref{torusbrane}) of the torus at the self-dual point. 
To do this, we have identified the `natural' $\mathbb{Z}_4$ symmetry 
(\ref{naturalZ_4}) on the torus with the quantum symmetry in the Gepner 
model, and the $\mathbb{Z}_4$ symmetry involving a T-duality 
transformation in both torus directions (\ref{g'action}) with the phase 
shift $e^{i\frac{\pi}{2}m_1}$. For a convenient choice of some relative 
phases, this has yielded an identification of angles, positions, and Wilson 
lines of the torus branes corresponding to the considered branes in the 
Gepner model in terms of the labels of the latter (Table 
\ref{identificationtable}).\\

The $N=2$ A-type boundary states on the torus can all be given by $U(1)$ 
branes, satisfying gluing conditions that involve (twice) the angle of 
the brane on the torus analogously to the electric flux of $U(1)$ branes 
in electric fields (\ref{torusgluingconditions}). Both $\mathbb{Z}_4$ 
symmetries rotate the angle by 90 degrees (or $\phi\mapsto\phi+\pi$). 
With our definitions, the values of position and Wilson line of a brane 
are kept fixed under the first symmetry, and exchanged under the second. 
Hence the first symmetry can be seen as a mere rotation of the brane around the 
point $(\pi,\pi)$ in the diagrams, leaving the distance to the origin
and the Wilson line fixed, while the second symmetry in general 
involves a shift in position.\\
The Gepner branes can be identified with matrix factorisations of a 
corresponding Landau-Ginzburg orbifold. Although the direct 
identification of Gepner brane labels with e.g. the phases of the 
orbifold matrices $\gamma$ is heavily depending on our conventions, 
we can for the considered model make the more general remark that a 
shift by $\tfrac{\pi}{2}$ in the phase of $\gamma$ corresponds to a 
rotation of 90 degrees of the corresponding brane on the torus, or a 
shift of $\phi$ by $\pi$, respectively.

\section*{Acknowledgements}
I would like to thank my advisor Matthias Gaberdiel for suggesting the 
problem, and for his constant advice and support. I am very grateful to 
Stefan Fredenhagen for his explanations and help. I would also like to 
thank Ilka Brunner, Eleonora Dell'Aquila, and Peter Kaste for helpful 
discussions. This research was supported by the Swiss National Science 
Foundation.

\appendix
\section{Conventions for the $N=2$ Minimal Models}
The $N=2$ algebra is generated by the modes $L_n$ of the energy-momentum 
tensor, the modes $J_n$ of the $U(1)$ current, and the modes $G_r^{\pm}$ 
of the two supercharges, where $n\in\mathbb{Z}$ and $r\in\mathbb{Z}$ for 
the R sector or $r\in\mathbb{Z}+\tfrac{1}{2}$ for the NS sector. They 
obey the (anti-)commutation relations
\begin{eqnarray}
[ L_m,\,L_n ] &=& (m-n)L_{m+n}+\tfrac{c}{12}m(m^2-1)\delta_{m,-n}\,,
\nonumber\\[2pt]
 [ L_m,\,J_n ] &=& -nJ_{m+n}\,,\nonumber\\[2pt]
 [ L_m,\,G_r^{\pm} ] &=& \left(\tfrac{m}{2}-r\right)G_{m+r}^{\pm}\,,
\nonumber\\[2pt]
 [ J_m,\,J_n ] &=& \tfrac{c}{3}m\delta_{m,-n}\,,\nonumber\\[2pt]
 [ J_m,\,G_r^{\pm} ] &=& \pm G_{m+r}^{\pm}\,,\nonumber\\[2pt]
 \{G_r^{+},\,G_s^{-}\} &=& 2L_{r+s}+(r-s)J_{r+s}+\tfrac{c}{3}
\left({r^2}-\tfrac{1}{4}\right)\delta_{r,-s}\,;\nonumber
\end{eqnarray}
all the other (anti-)commutators vanish.
The $N=2$ minimal models at level $k\in\mathbb{N}$ have central charge
\begin{equation}
c=\frac{3k}{k+2}
\end{equation}
and are described by means of the coset construction
\begin{equation}
\frac{su(2)_k\otimes u(1)_4}{u(1)_{2(k+2)}}\,.
\end{equation}
Highest weight representations of the coset construction are labelled by
\begin{equation}
l\in\{0,\ldots,k\}\,,\quad m\in\mathbb{Z}_{2(k+2)}\,,\quad s\in\mathbb{Z}_4\,,
\end{equation}
with the selection rule that $l+m+s$ must be even.
The corresponding highest weight state is denoted as
\begin{equation}
\Phi^{l,s}_{m}\equiv |l,\,m,\,s\rangle\,,
\end{equation}
with highest weight and charge
\begin{equation}
h=\frac{l(l+2)-m^2}{4(k+2)}+\frac{s^2}{8}\;\mathrm{mod}\,1\,,\qquad 
q=\frac{s}{2}-\frac{m}{k+2}\;\mathrm{mod}\,2\,.
\end{equation}
The set of labels $(k-l,\,m+k+2,\,s+2)$ gives a representation identical 
to the one with labels $(l,\,m,\,s)$; we denote the equivalence class by 
$[l,\,m,\,s]$. A complete $N=2$ NS representation is given by the direct 
sum $(l,\,m,\,0)\oplus(l,\,m,\,2)$, a complete R representation is 
$(l,\,m,\,1)\oplus(l,\,m,\,3)$, where one part of the direct sum contains 
the states at even and at odd fermion number respectively. The chiral 
primaries are given by the labels $(l,\,l,\,0)$ or $(l,\,-l-2,\,2)$ in 
the NS sector and by $(l,\,l+1,\,1)$ or $(l,\,-l-1,\,-1)$ in the R sector.
The modular $S$ matrix of the coset theory is
\begin{equation}
S_{LMS,lms}=\frac{1}{\sqrt{2(k+2)}}S_{Ll}e^{i\frac{\pi}{k+2}Mm}
e^{-i\frac{\pi}{2}Ss}\,,
\end{equation}
where
\begin{equation}
S_{Ll}=\sqrt{\frac{2}{k+2}}\sin\left(\frac{\pi}{k+2}(L+1)(l+1)\right)
\end{equation}
is the modular $S$ matrix of $su(2)_k$. The spectral flow of unit 
$\tfrac{1}{2}$ acts on the coset labels by fusion with $(0,\,1,\,1)$.

\section{Table for the Gepner model and its relation to the torus}

A list of the minimal model representations of the $N=2$ superconformal 
algebra at $k=2$ in terms of the coset labels is given in Table 5.
If one follows the comparison of the characters that lead
(\ref{decomposition}) to higher orders, one is lead to the 
following general formula for diagonal representations at $c=3$ contained 
in the tensor product of two minimal models at $k=2$:
\begin{table}\label{minmodlist}
\begin{center}
\begin{tabular}{|c|c|c||c|c||c|c|c||c|c|}
\hline\multicolumn{5}{|c||}{}&\multicolumn{5}{c|}{}\\[-11pt]
\multicolumn{5}{|c||}{NS -- sector}&\multicolumn{5}{c|}{R -- sector}\\[4pt]
\hline&&&&&&&&&\\[-11pt]
$l$&$m$&$s$&$h$&$q$&$l$&$m$&$s$&$h$&$q$\\[4pt]
\hline&&&&&&&&&\\[-11pt]
$\mathbf{0}$ & $\mathbf{0}$ & $\mathbf{0}$ & $\mathbf{0}$ & 
$\mathbf{0}$                               & $\mathbf{0}$ & 
$\mathbf{1}$ & $\mathbf{1}$ & $\mathbf{\tfrac{1}{16}}$ & 
$\mathbf{\tfrac{1}{4}}$  \\[4pt]
0 & 2 & 2 & $\tfrac{1}{4}$ & $\tfrac{1}{2}$ 	  & 0 & 3 & 
3 & $\tfrac{9}{16}$ & $\tfrac{3}{4}$ \\[4pt]
0 & 4 & 2 & $\tfrac{1}{2}$ & 0                  & 0 & 5 & 3 
& $\tfrac{9}{16}$ & $\tfrac{1}{4}$  \\[4pt]
$\mathbf{0}$ & $\mathbf{6}$ & $\mathbf{2}$ & $\mathbf{\tfrac{1}{4}}$ 
& $\mathbf{-\tfrac{1}{2}}$ 	  & $\mathbf{0}$ & $\mathbf{7}$ & 
$\mathbf{3}$ & $\mathbf{\tfrac{1}{16}}$ & $\mathbf{-\tfrac{1}{4}}$ 
\\[4pt]
$\mathbf{1}$ & $\mathbf{1}$ & $\mathbf{0}$ & $\mathbf{\tfrac{1}{8}}$ 
& $\mathbf{-\tfrac{1}{4}}$ 	  & 1 & 0 & 1 & $\tfrac{5}{16}$ & 
$\tfrac{1}{2}$  \\[4pt]
1 & 3 & 2 & $\tfrac{1}{8}$ & $\tfrac{1}{4}$ 	  & $\mathbf{1}$ 
& $\mathbf{2}$ & $\mathbf{1}$ & $\mathbf{\tfrac{1}{16}}$ & 
$\mathbf{0}$             \\[4pt]
\hline
\end{tabular}
\caption{\emph{List of coset labels, highest weights and charges 
for the representations of the $N=2$ algebra at level $k=2$. For 
the NS sector, the coset label $s$ indicates the bosonic subalgebra 
of even fermion number, in the R sector the coset labels give the 
subalgebra of the highest weight state which is annihilated by 
$G_0^+$. Bold face indicates chiral primaries.}}
\end{center}
\end{table}
\begin{eqnarray}
(0,\,0)\otimes(0,\,0)&=&(0,\,0)\displaystyle{
\bigoplus_{m\in\mathbb{Z}}}\left(\tfrac{8|m|-1}{2},\,
\mathrm{sign}(\tfrac{8m-1}{2})\right)
\displaystyle{\bigoplus_{n\in\mathbb{N}}}\left(n^2,\,0\right)
\nonumber\\
&&\phantom{(0,\,0)}
\displaystyle{\bigoplus_{p, q\in\mathbb{N}}}\left(p^2+q^2,\,0\right)
\,,\nonumber\\
\left(\tfrac{1}{2},\,0\right)\otimes(0,\,0)&=&
\displaystyle{\bigoplus_{n\,\mathrm{odd}}}
\left(\tfrac{n^2}{2},\,0\right)
\displaystyle{\bigoplus_{p^2+q^2\,\mathrm{odd}}}
\left(\tfrac{p^2+q^2}{2},\,0\right)\,,\nonumber\\
\left(\tfrac{1}{8},\,\tfrac{1}{4}\right)\otimes
\left(\tfrac{1}{8},\,-\tfrac{1}{4}\right)&=&
\displaystyle{\bigoplus_{n\,\mathrm{odd}}}
\left(\tfrac{n^2}{4},\,0\right)
\displaystyle{\bigoplus_{p^2+q^2\,\mathrm{odd}}}
\left(\tfrac{p^2+q^2}{4},\,0\right)\,.\nonumber
\end{eqnarray}
Here, $m$ runs over all integers, whereas $n,\,p,\,$ and $q$ only 
take values in the set of natural numbers. By applying the spectral 
flow on both factors on the left hand side as well as on the summands 
on the right-hand side, one obtains the formulae for the other tensor 
products appearing in the Gepner model. This formula has not been proved,
but it has been checked numerically up to level 50.


\begin{thebibliography}{99}

\bibitem{Brunner:1999jq}
  I.~Brunner, M.~R.~Douglas, A.~E.~Lawrence and C.~Romelsberger,
  \textit{D-branes on the quintic}, 
  JHEP {\bf 0008} (2000) 015
  [arXiv:hep-th/9906200].
  %%CITATION = HEP-TH 9906200;%%

\bibitem{Hori:2000ck}
  K.~Hori, A.~Iqbal and C.~Vafa,
  \textit{D-branes and mirror symmetry},
  arXiv:hep-th/0005247.
  %%CITATION = HEP-TH 0005247;%%

\bibitem{Diaconescu:2000ec}
  D.~E.~Diaconescu and M.~R.~Douglas,
  \textit{D-branes on stringy Calabi-Yau manifolds},
  arXiv:hep-th/0006224.
  %%CITATION = HEP-TH 0006224;%%

\bibitem{Mayr:2000as}
  P.~Mayr,
  \textit{Phases of supersymmetric D-branes on Kaehler manifolds and
    the McKay correspondence}, 
  JHEP {\bf 0101}, 018 (2001)
  [arXiv:hep-th/0010223].
  %%CITATION = HEP-TH 0010223;%%

\bibitem{Tomasiello:2000ym}
  A.~Tomasiello,
  \textit{D-branes on Calabi-Yau manifolds and helices},
  JHEP {\bf 0102}, 008 (2001)
  [arXiv:hep-th/0010217].
  %%CITATION = HEP-TH 0010217;%%

\bibitem{Govindarajan:2001jc}
  S.~Govindarajan and T.~Jayaraman,
  \textit{D-branes and vector bundles on Calabi-Yau manifolds: A view
    from the helix},  
  arXiv:hep-th/0105216.
  %%CITATION = HEP-TH 0105216;%%

\bibitem{Aspinwall:2006ib}
  P.~S.~Aspinwall,
  \textit{The Landau-Ginzburg to Calabi-Yau dictionary for D-branes},
  arXiv:hep-th/0610209.
  %%CITATION = HEP-TH 0610209;%%

\bibitem{Recknagel:1997sb}
  A.~Recknagel and V.~Schomerus,
  \textit{D-branes in Gepner models}, 
  Nucl.\ Phys.\ B {\bf 531} (1998) 185
  [arXiv:hep-th/9712186].
  %%CITATION = HEP-TH 9712186;%%

\bibitem{Ooguri:1996ck}
  H.~Ooguri, Y.~Oz and Z.~Yin,
  \textit{D-branes on Calabi-Yau spaces and their mirrors}, 
  Nucl.\ Phys.\ B {\bf 477} (1996) 407
  [arXiv:hep-th/9606112].
  %%CITATION = HEP-TH 9606112;%%

\bibitem{Govindarajan:2000ef}
  S.~Govindarajan, T.~Jayaraman and T.~Sarkar,
  \textit{On D-branes from gauged linear sigma models},
  Nucl.\ Phys.\ B {\bf 593}, 155 (2001)
  [arXiv:hep-th/0007075].
  %%CITATION = HEP-TH 0007075;%%

\bibitem{Kapustin:2002bi}
  A.~Kapustin and Y.~Li,
  \textit{D-branes in Landau-Ginzburg models and algebraic geometry}, 
  JHEP {\bf 0312} (2003) 005
  [arXiv:hep-th/0210296].
  %%CITATION = HEP-TH 0210296;%%

\bibitem{Brunner:2003dc}
  I.~Brunner, M.~Herbst, W.~Lerche and B.~Scheuner,
  \textit{Landau-Ginzburg realization of open string TFT}, 
  arXiv:hep-th/0305133.
  %%CITATION = HEP-TH 0305133;%%

\bibitem{Hori:2004zd}
  K.~Hori and J.~Walcher,
  \textit{D-branes from matrix factorizations}, 
  Comptes Rendus Physique {\bf 5} (2004) 1061
  [arXiv:hep-th/0409204].
  %%CITATION = HEP-TH 0409204;%%

\bibitem{Warner:1995ay}
  N.~P.~Warner,
  \textit{Supersymmetry in boundary integrable models},
  Nucl.\ Phys.\ B {\bf 450} (1995) 663
  [arXiv:hep-th/9506064].
  %%CITATION = HEP-TH 9506064;%%

\bibitem{Lerche:1989cs}
  W.~Lerche, D.~Lust and N.~P.~Warner,
  \textit{Duality symmetries in N=2 Landau-Ginzburg models},
  Phys.\ Lett.\ B {\bf 231} (1989) 417.
  %%CITATION = PHLTA,B231,417;%%

\bibitem{Recknagel:2002qq}
  A.~Recknagel,
  \textit{Permutation branes}, 
  JHEP {\bf 0304} (2003) 041
  [arXiv:hep-th/0208119].
  %%CITATION = HEP-TH 0208119;%%

\bibitem{Gaberdiel:2002jr}
  M.~R.~Gaberdiel and S.~Sch\"afer-Nameki,
  \textit{D-branes in an asymmetric orbifold},
  Nucl.\ Phys.\ B {\bf 654} (2003) 177
  [arXiv:hep-th/0210137].
  %%CITATION = HEP-TH 0210137;%%

\bibitem{Chun:1991js}
  E.~J.~Chun, J.~Lauer and H.~P.~Nilles,
  \textit{Equivalence of Z(N) orbifolds and Landau-Ginzburg models},
  Int.\ J.\ Mod.\ Phys.\ A {\bf 7} (1992) 2175.
  %%CITATION = IMPAE,A7,2175;%%

\bibitem{Brunner:2005fv}
  I.~Brunner and M.~R.~Gaberdiel,
  \textit{Matrix factorisations and permutation branes}, 
  JHEP {\bf 0507} (2005) 012
  [arXiv:hep-th/0503207].
  %%CITATION = HEP-TH 0503207;%%

\bibitem{Brunner:2005pq}
  I.~Brunner and M.~R.~Gaberdiel,
  \textit{The matrix factorisations of the D-model}, 
  J.\ Phys.\ A {\bf 38} (2005) 7901
  [arXiv:hep-th/0506208].
  %%CITATION = HEP-TH 0506208;%%

\bibitem{Gutperle:1998hb}
  M.~Gutperle and Y.~Satoh,
  \textit{D-branes in Gepner models and supersymmetry}, 
  Nucl.\ Phys.\ B {\bf 543} (1999) 73
  [arXiv:hep-th/9808080].
  %%CITATION = HEP-TH 9808080;%%

\bibitem{Govindarajan:1999js}
  S.~Govindarajan, T.~Jayaraman and T.~Sarkar,
  \textit{Worldsheet approaches to D-branes on supersymmetric cycles}, 
  Nucl.\ Phys.\ B {\bf 580} (2000) 519
  [arXiv:hep-th/9907131].
  %%CITATION = HEP-TH 9907131;%%

\bibitem{Dell'Aquila:2005jg}
  E.~Dell'Aquila,
  \textit{D-branes in toroidal orbifolds and mirror symmetry},
  JHEP {\bf 0604} (2006) 035
  [arXiv:hep-th/0512051].
  %%CITATION = HEP-TH 0512051;%%

\bibitem{Gaberdiel:2004nv}
  M.~R.~Gaberdiel and H.~Klemm,
  \textit{N = 2 superconformal boundary states for free bosons and fermions}, 
  Nucl.\ Phys.\ B {\bf 693} (2004) 281
  [arXiv:hep-th/0404062].
  %%CITATION = HEP-TH 0404062;%%

\bibitem{Ishibashi:1988kg}
  N.~Ishibashi,
  \textit{The boundary and crosscap states in conformal field
    theories}, 
  Mod.\ Phys.\ Lett.\ A {\bf 4} (1989) 251.
  %%CITATION = MPLAE,A4,251;%%

\bibitem{Klemm:2005dt}
  H.~Klemm,
  \textit{$N=2$ Superconformal boundary states}, 
  Doctoral Thesis, Department of Mathematics, King's College,
  University of London (2005).

\bibitem{Wendland:2005nz}
  K.~Wendland,
  \textit{A family of SCFTs hosting all 'very attractive' relatives of
    the (2)**4 Gepner model}, 
  JHEP {\bf 0603} (2006) 102
  [arXiv:hep-th/0512223].
  %%CITATION = HEP-TH 0512223;%%

\bibitem{Gepner:1987vz}
  D.~Gepner,
  \textit{Exactly solvable string compactifications on manifolds of
    SU(N) holonomy}, 
  Phys.\ Lett.\ B {\bf 199} (1987) 380.
  %%CITATION = PHLTA,B199,380;%%

\bibitem{Gepner:1987qi}
  D.~Gepner,
  \textit{Space-time supersymmetry in compactified string theory and
    superconformal models}, 
  Nucl.\ Phys.\ B {\bf 296} (1988) 757.
  %%CITATION = NUPHA,B296,757;%%

\bibitem{Schellekens:1989am}
  A.~N.~Schellekens and S.~Yankielowicz,
  \textit{Extended chiral algebras and modular invariant partition
    functions}, 
  Nucl.\ Phys.\ B {\bf 327} (1989) 673.
  %%CITATION = NUPHA,B327,673;%%

\bibitem{Fuchs:2000fd}
  J.~Fuchs, P.~Kaste, W.~Lerche, C.~A.~Lutken, C.~Schweigert and J.~Walcher,
  \textit{Boundary fixed points, enhanced gauge symmetry and singular
    bundles on K3},  
  Nucl.\ Phys.\ B {\bf 598} (2001) 57
  [arXiv:hep-th/0007145].
  %%CITATION = HEP-TH 0007145;%%

\bibitem{Fuchs:2000gv}
  J.~Fuchs, C.~Schweigert and J.~Walcher,
  \textit{Projections in string theory and boundary states for Gepner models}, 
  Nucl.\ Phys.\ B {\bf 588} (2000) 110
  [arXiv:hep-th/0003298].
  %%CITATION = HEP-TH 0003298;%%

\bibitem{Fredenhagen:2006qw}
  S.~Fredenhagen and M.~R.~Gaberdiel,
  \textit{Generalised N = 2 permutation branes},
  JHEP {\bf 0611} (2006) 041
  [arXiv:hep-th/0607095].
  %%CITATION = HEP-TH 0607095;%%

\bibitem{Brunner:2004zd}
  I.~Brunner, K.~Hori, K.~Hosomichi and J.~Walcher,
  \textit{Orientifolds of Gepner models},
  arXiv:hep-th/0401137.
  %%CITATION = HEP-TH 0401137;%%

\bibitem{Kapustin:2003ga}
  A.~Kapustin and Y.~Li,
  \textit{Topological correlators in Landau-Ginzburg models with boundaries}, 
  Adv.\ Theor.\ Math.\ Phys.\  {\bf 7} (2004) 727
  [arXiv:hep-th/0305136].
  %%CITATION = HEP-TH 0305136;%%

\bibitem{Walcher:2004tx}
  J.~Walcher,
  \textit{Stability of Landau-Ginzburg branes},
  J.\ Math.\ Phys.\  {\bf 46} (2005) 082305
  [arXiv:hep-th/0412274].
  %%CITATION = HEP-TH 0412274;%%

\bibitem{Caviezel:2005th}
  C.~Caviezel, S.~Fredenhagen and M.~R.~Gaberdiel,
  \textit{The RR charges of A-type Gepner models},
  JHEP {\bf 0601} (2006) 111
  [arXiv:hep-th/0511078].
  %%CITATION = HEP-TH 0511078;%%

\end{thebibliography}
\end{document}